\documentclass[11pt,letterpaper,twocolumn]{aastex63}
\usepackage{verbatim}
\usepackage{amsmath}
\usepackage{epsfig}
\usepackage{amsmath}
\usepackage{natbib}
\usepackage{mathtools}

\begin{document}
 
\title{A Decade of Linear and Circular Polarimetry with the POLISH2 Polarimeter}
\author{Sloane J. Wiktorowicz}
\affiliation{Remote Sensing Department, The Aerospace Corporation, El Segundo, CA 90245}
\email{s.wiktorowicz@aero.org}

\author{Agnieszka S\l{}owikowska}
\affiliation{Institute of Astronomy, Faculty of Physics, Astronomy and Informatics, Nicolaus Copernicus University in Toru\'n, Grudziadzka 5, PL-87-100, Toru\'n, Poland}

\author{Larissa A. Nofi}
\affiliation{Remote Sensing Department, The Aerospace Corporation, El Segundo, CA 90245}

\author{Nicole Rider}
\affiliation{Department of Physics and Astronomy, University of North Carolina, Chapel Hill, NC, 27599}

\author{Angie Wolfgang}
\affiliation{SiteZeus, Tampa, FL 33602}

\author{Ninos Hermis}
\affiliation{Faculty of Physics, Ludwig-Maximilians-Universit\"{a}t, Munich, Germany}

\author{Daniel Jontof-Hutter}
\affiliation{Department of Physics, University of the Pacific, Stockton, CA 95211}

\author{Amanda J. Bayless}
\affiliation{Remote Sensing Department, The Aerospace Corporation, El Segundo, CA 90245}

\author{Gary M. Cole}
\affiliation{Starphysics Observatory, Reno, NV 89511}

\author{Kirk B. Crawford}
\affiliation{Sensors \& Systems Department, The Aerospace Corporation, El Segundo, CA 90245}

\author{Valeri V. Tsarev}
\affiliation{Sensors \& Systems Department, The Aerospace Corporation, El Segundo, CA 90245}

\author{Michael C. Owens}
\affiliation{Remote Sensing Department, The Aerospace Corporation, El Segundo, CA 90245}

\author{Ernest G. Jaramillo}
\affiliation{Remote Sensing Department, The Aerospace Corporation, El Segundo, CA 90245}

\author{Geoffrey A. Maul}
\affiliation{Sensors \& Systems Department, The Aerospace Corporation, El Segundo, CA 90245}

\author{James R. Graham}
\affiliation{Astronomy Department, University of California, Berkeley, CA 94720}

\author{Maxwell A. Millar-Blanchaer}
\affiliation{Physics Department, University of California, Santa Barbara, CA 93106}

\author{Kimberly Bott}
\affiliation{Earth and Planetary Sciences Department, University of California, Riverside, CA 92521}

\author{Jon C. Mauerhan}
\affiliation{Remote Sensing Department, The Aerospace Corporation, El Segundo, CA 90245}

\date{\today}

\begin{abstract}

The POLISH2 optical polarimeter has been in operation at the Lick Observatory 3-m Shane telescope since 2011, and it was commissioned at the Gemini North 8-m in 2016. This instrument primarily targets exoplanets, asteroids, and the Crab pulsar, but it has also been used for a wide variety of planetary, galactic, and supernova science. POLISH2's photoelastic modulators, employed instead of rotating waveplates or ferro-electric liquid crystal modulators, offer the unprecedented ability to achieve sensitivity and accuracy of order 1 ppm (0.0001\%), which are difficult to obtain with conventional polarimeters. Additionally, POLISH2 simultaneously measures intensity (Stokes $I$), linear polarization (Stokes $Q$ and $U$), and circular polarization (Stokes $V$), which fully describe the polarization state of incident light. We document our laboratory and on-sky calibration methodology, our archival on-sky database, and we demonstrate conclusive detection of circular polarization of certain objects. \\

\end{abstract}
 
\section{Introduction}

Reflected or scattered light from astronomical objects imprints spectroscopic and polarimetric signatures that are crucial to the interpretation of those objects. Calibration procedures, both in the laboratory and on-sky, allow quantitative assessment of the sensitivity and accuracy of instrumentation, which is critical to science. Here, we define ``sensitivity" as the ability to measure a change, which is fundamentally limited by photon noise. We define ``accuracy" as the ability to measure a change intrinsic to the object of the experiment, rather than to the telescope and instruments used to obtain data. Accuracy is rarely limited by photon noise.

While spectroscopic calibration tends to be relatively straightforward in principle, involving standard, man-made or astronomical light sources with known spectral features, polarimetric calibration can be quite difficult. Since reflection alters the polarization state of incident light, it can be difficult to inject a source of known polarization into an instrument in the laboratory. For example, laboratory calibration of the Gemini Planet Imager polarization mode \citep{Wiktorowicz2012} suggests intrinsic linear and circular polarization of the instrument to be two to three times larger than that found during on-sky commissioning \citep{Wiktorowicz2014}. This may be due to reflection-induced conversion of linear to circular polarization in the laboratory setup, and it is a nearly unavoidable consequence of injecting light from a nearby, laboratory source into an instrument designed to mount to a telescope focusing at infinity.

A variety of laboratory polarization optics may be used to convert lamp light, assumed to be unpolarized, into pure linear or circular polarization. However, high accuracy calibration requires astronomical objects due to 1) their long-term stability, 2) the ability to compare measurements of the same object with the literature, and 3) the availability of a large dynamic range of partially polarized calibration objects. The latter point is crucial, as laboratory calibration using 100\% polarized sources may not necessarily enable calibration to the 1 ppm level.

In Section \ref{sec:POLISH2}, we describe the POLISH2 instrument and its general calibration methodology using a combination of laboratory and astronomical sources. We discuss specific laboratory and on-sky calibration results in Section \ref{sec:labcal}, and we tabulate the POLISH2 database in Section \ref{sec:database}.
 
\section{The POLISH2 Instrument}
\label{sec:POLISH2}
\subsection{Design}
\subsubsection{Optics}
 
\begin{figure}
\centering
\includegraphics[scale=0.049]{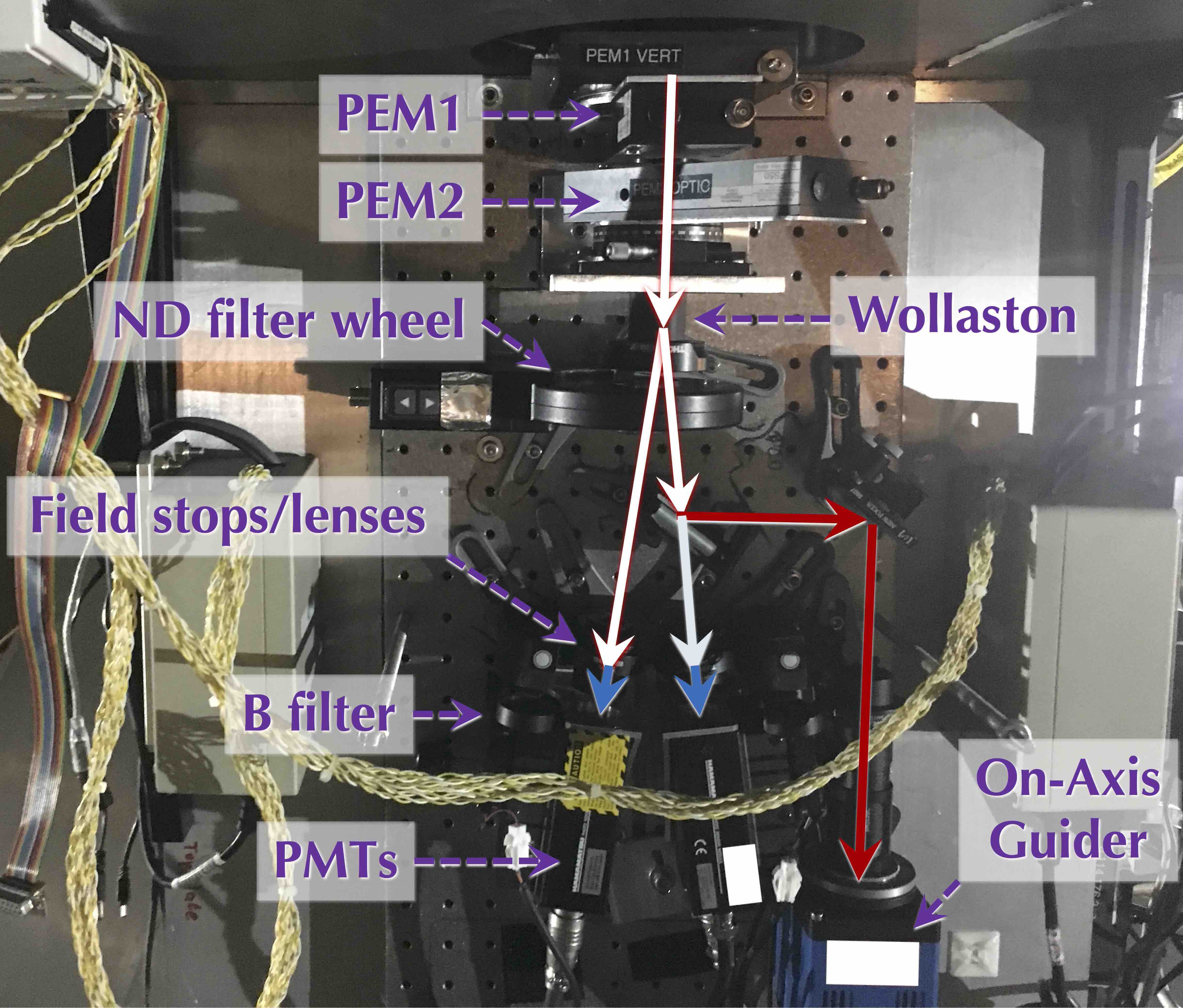}
\caption{POLISH2 optical layout.}
\label{fig:layout}
\end{figure}

POLISH2 \citep{WiktorowiczNofi2015} is an aperture integrated, optical polarimeter with two Hinds Instruments I/FS40 and I/FS50 photoelastic modulators (PEMs) and a two-wedge, Karl Lambrecht calcite Wollaston prism (polarization beamsplitter) directly downstream of the telescope secondary mirror. POLISH2 has operated at three telescopes with straight Cassegrain foci: the Gemini North 8-m ($f/16$), the Lick 3-m ($f/17$), and the Lick 1-m ($f/17$). The first PEM is oriented with its compression/extension axis perpendicular to the Wollaston prism deviation axis, while the second PEM is oriented $45^\circ$ clockwise from the first PEM when looking into the telescope secondary. The PEMs are followed by the Wollaston prism, which bifurcates the beam and doubles the throughput with respect to a static linear polarizer analyzer. Alterations to the polarization state of light downstream of the Wollaston do not bias polarization measurements, as the PEM/Wollaston combination converts incident polarization into intensity variations. Next, both Wollaston output beams pass through the same aperture of a Thorlabs FW102C six-position filter wheel, which enables broadband or neutral density filters to be inserted into the beam. A Semrock dichroic is located in one of the two Wollaston output beams, and it directs light redward of 750 nm to an f/3 focal reducer and PCO Edge 4.2 sCMOS 2k x 2k guide camera for on-axis guiding (Figure \ref{fig:layout}).

\begin{figure}
\centering
\includegraphics[scale=0.34]{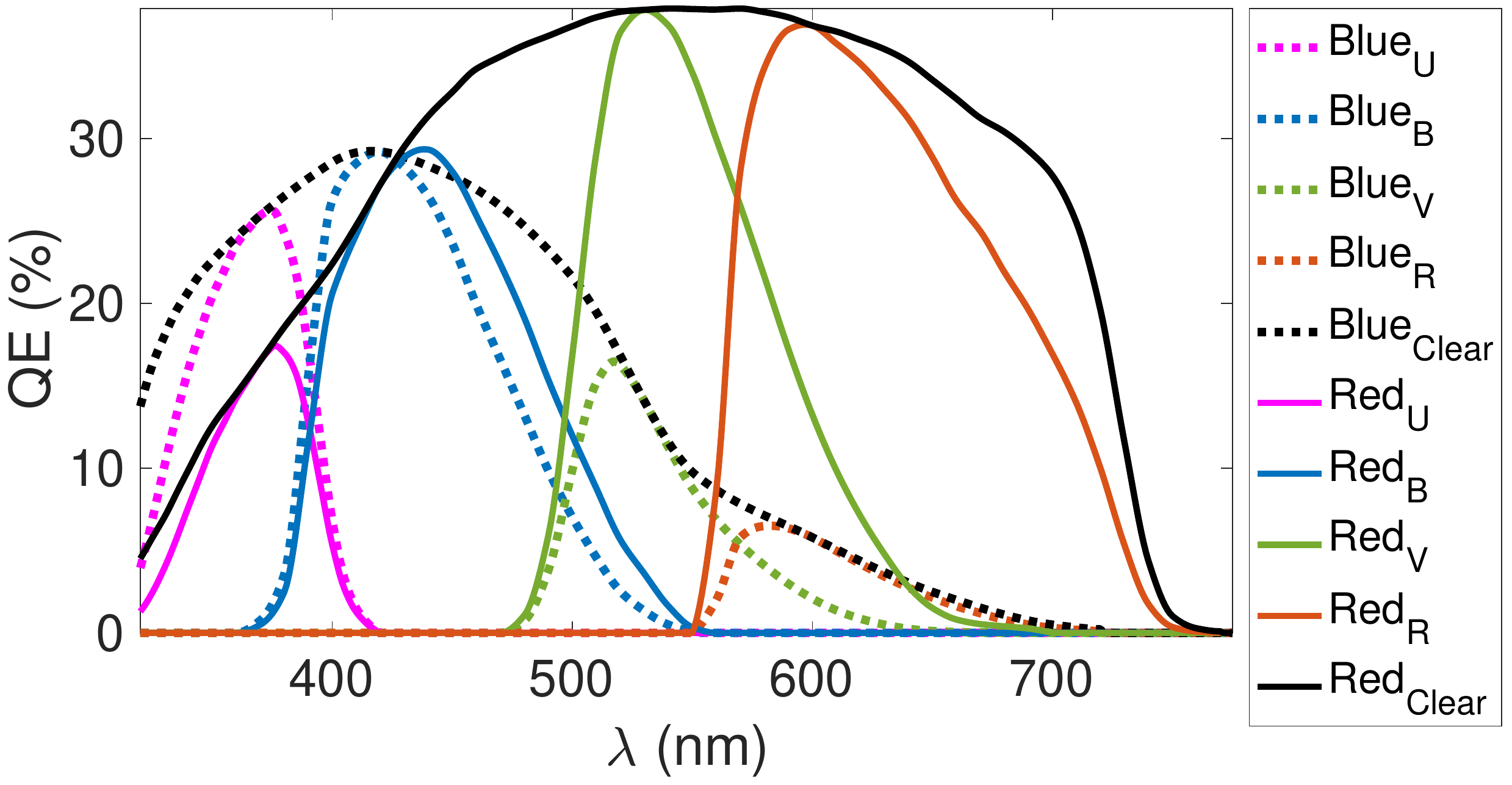}
\caption{POLISH2 QE curves for the blue-sensitive Hamamatsu H10721-210 SEL (dotted curves) and the red-sensitive H7422P-40 SEL PMTs (solid curves). These are convolved with Bessell $UBVR$ bandpasses and are also shown with clear, unfiltered bandpasses. The $R$ bandpasses are illustrative, as POLISH2 has not obtained $R$ band measurements to date.}
\label{fig:qe}
\end{figure}

 The Wollaston causes the telescope secondary mirror to focus onto a pair of 5 mm diameter field stops, which correspond to an FOV of $8\arcsec$ diameter at Gemini North, $19\arcsec$ at the Lick 3-m, and $58\arcsec$ at the Lick 1-m. While smaller diameter field stops would reduce sky background contamination, vignetting due to lateral chromatism of the Wollaston limits the size of the field stops. Originally, field lenses then imaged the telescope pupil onto a pair of blue-sensitive Hamamatsu H10721-110 SEL photomultiplier tube (PMT) science detectors. Blue sensitivity is desired to search for polarized Rayleigh scattering from short-period exoplanets. These PMTs were upgraded to Hamamatsu H10721-210 SEL for the August 2018 Gemini North run, which increased QE but maintained the shape of the QE curve. Red-sensitive Hamamatsu H7422P-40 SEL PMTs were commissioned during the September 2017 Lick 3-m for solar system asteroid and Crab Pulsar science. Only one set of PMTs is used on a given observing run. Figure \ref{fig:qe} and Table \ref{tab:qe} illustrate the effective bandpasses of the post-2017 POLISH2 PMTs with one airmass of extinction taken into account. In Table \ref{tab:qe}, wavelength of peak QE is listed second and is bracketed by wavelengths of half-maximum QE. Values are sorted by the red wavelength of half-maximum QE.
 
\begin{deluxetable}{cccccccccc}
\tabletypesize{\normalsize}
\tablecaption{POLISH2 Effective Bandpasses}
\tablewidth{0pt}
\tablehead{
\colhead{PMT} & \colhead{Band} & \colhead{PMT Bandpass (nm)}}
\startdata
Blue		& $U$	& 335,   375,   394 \\
$\cdots$	& $B$	& 390,   419,   475 \\
$\cdots$	& Clear	& 322,   416,   527 \\
$\cdots$	& $V$	& 498,   518,   552 \\
$\cdots$	& $R$	& 564,   583,   632 \\
\hline
Red		& $U$	& 343,   376,   395 \\
$\cdots$	& $B$	& 394,   438,   491 \\
$\cdots$	& $V$	& 502,   530,   586 \\
$\cdots$	& $R$	& 565,   597,   694 \\
$\cdots$	& Clear	& 383,   542,   720
\label{tab:qe}
\enddata
\end{deluxetable}
 
 While the blue- and red-sensitive PMTs have similar $B$ band QE, the blue-sensitive PMTs are utilized during exoplanet Rayleigh scattering runs due to their 50x higher saturation limits. This enables the blue-sensitive PMTs to operate on stars over 4 mag brighter than the red-sensitive PMTs without the use of ND filters. This is crucial in collecting the high photon rates necessary to reach part-per-million sensitivity on both exoplanet host stars and bright, weakly polarized calibrator stars. On the other hand, the red-sensitive PMTs are used for asteroid and Crab pulsar science to maximize throughput. A pair of Thorlabs MFF101 motorized flip mounts enable a Bessell $B$-band filter to be inserted into each Wollaston beam to search for Rayleigh scattering from close-in exoplanets. These free the upstream filter wheel to house ND filters.
 
 As even the red-sensitive PMTs operate below 750 nm, the guider arm's 750 nm short-pass dichroic imparts essentially no bias to either the polarization measurement or the intensity on the PMTs. Even though the PCO guider has a fairly blue bandpass but is fed with light above 750 nm, the guider is able to detect stars too faint for the POLISH2 PMTs. This is because the continuous readout of the PMTs enables operation at the high photon rates necessary to reach 1 ppm sensitivity on exoplanet host stars, but it does not allow integration of light from faint targets like conventional pixels. At the bright end, Venus is not an atypical target for Lick 3-m POLISH2, while the Crab pulsar is the faintest target observed so far. Thus, POLISH2 operates with a photon-limited dynamic range of $\sim 20$ magnitudes. 

\subsubsection{Data Acquisition}
\label{sec:dataacq}

POLISH2's PEMs \citep{Kemp1969} are bars of fused silica with resonant frequencies of 40 and 50 kHz. When powered on, the PEMs are piezoelectrically forced at these frequencies, and light passing through them experiences a sinusoidal temporal modulation in retardance. This effectively converts incident linear and circular polarization to and from each other. The Wollaston then converts this time-variable polarization state into a time-variable intensity that is measured by the PMTs. While a linear polarizer would also perform this conversion, it would discard half of the intensity of the beam and reduce SNR. The use of a Wollaston prism and a pair of PMTs enables instrumental throughput to be $70-75\%$, which maximizes photon rates and enables ppm-level sensitivity on bright targets. The orientation of the two PEMs and Wollaston enable simultaneous measurement of intensity (Stokes $I$), linear polarization (Stokes $Q$ and $U$), and circular polarization (Stokes $V$). Each Stokes parameter is therefore converted to intensity variations at specific frequencies (AC coupled modulation), which are composed of linear combinations of the 40 and 50 kHz resonant frequencies of the PEMs. That is, Stokes $Q$ is modulated primarily at $2 \times 50 = 100$ kHz (even harmonics of the 50 kHz PEM), Stokes $U$ at $50 \pm 40 = 10$ and 90 kHz, Stokes $V$ at 50 kHz (odd harmonics of the 50 kHz PEM), and Stokes $I$ is derived from the time-averaged intensity.

\begin{figure}
\centering
\includegraphics[width=0.47\textwidth]{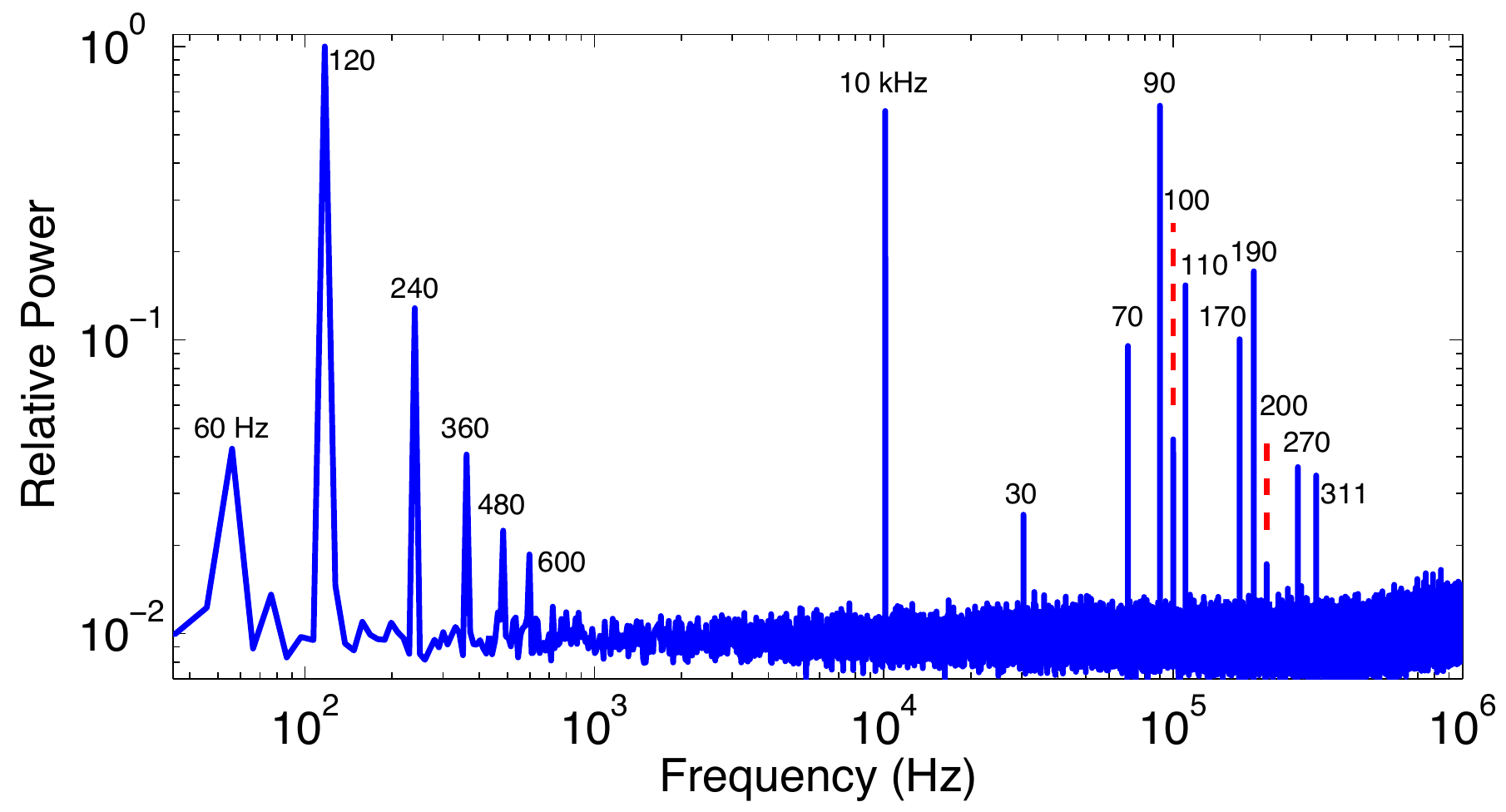}
\caption{FFT of raw Lick 3-m POLISH2 observations of incandescent flat field lamp light reflected by the inside of the closed dome. Frequencies of significant FFT peaks are labeled.  The presence of incident Stokes $Q$ is exposed by the power at 100 and 200 kHz (dashed, red lines), while all other harmonics at and above 10 kHz indicate Stokes $U$. Peaks below 1 kHz are due to intrinsic flicker of the lamp and dimmer circuit due to the United States power grid.}
\label{fig:fft}
\end{figure}

Figure \ref{fig:fft} shows the results of an FFT performed on raw POLISH2 signal obtained on an incandescent flat field lamp reflected by the closed Lick 3-m dome. Peak power near 120 Hz is indicative of AC power modulated at twice the frequency of voltage modulation, which is 60 Hz in the United States. Higher harmonics up to 1 kHz are generated by the TRIAC circuit that controls the lamp dimmer. In addition to lamp flicker at frequencies less than 1 kHz, modulation of intrinsic Stokes $Q$ and $U$ may be observed at frequencies of 10 kHz and higher. The differences in power between harmonics for the same Stokes parameters, e.g., 100 and 200 kHz for Stokes $Q$, are caused by variations in modulation efficiency for each harmonic and are corrected for in section \ref{sec:step1}. The difference in power between Stokes $Q$ and $U$ indicates the rotational orientation of the polarization vector scattered by the dome in the instrument frame.
 
A pair of Stanford Research SR570 transimpedance amplifiers (TIAs) convert the output current of each PMT to voltage with gain. While many TIA gains may be selected from, each has its own cutoff frequency that would affect the throughput at each of the harmonics in Figure \ref{fig:fft}. Rather than accept calibration offsets when combining measurements of the same target made at different gains, such as for observations of the same stars with the Lick 3-m and 1-m, we  typically set TIA gains to $5 \times 10^3$ or $10^5$ V/A for bright and faint targets, respectively.  As the TIA gain is in series with the PMT gain of up to $2 \times 10^6$ A/A, this combination of gains is key to attaining the large, $\sim 20$ magnitude dynamic range described above.

The voltage outputs of the TIAs are digitized at a sampling rate of 2 MHz and 16 bits per sample by a National Instruments PXIe-6124 data acquisition card (DAQ) in a PXIe-1062Q chassis. Simultaneously, the 40 and 50 kHz reference square waves from the PEMs are digitized, which enables synchronous demodulation of Stokes parameters via a lock-in amplifier algorithm in software. Thus, four simultaneous data channels are recorded to external hard drives with 16 bits per sample and a 2 MHz sampling rate, which explains the 1 MHz maximum frequency in the Figure \ref{fig:fft} FFT. On a long winter's night, raw data volume can surpass 500 GB. While the nightly volume of reduced data is typically 1,000 times smaller than that of raw data, raw data are archived in order to enable re-reduction as improvements are made to the data reduction pipeline. The size of the POLISH2 data archive is currently over 130 TB.

Since modulation harmonics $f_i$ (10, 50, 90, and 100 kHz) are linear combinations of PEM resonant frequencies $f_j$ (40 and 50 kHz), the lock-in amplifier data reduction code identifies the resonant frequency and phase $T_j$ of the PEM reference square waves. It then constructs in-phase $x_i (t)$ and quadrature $y_i (t)$ reference sinusoids for each modulation harmonic $i$ (10, 50, 90, and 100 kHz):
\begin{subequations}
\begin{align}
f_i & = m_i f_\text{PEM1} + n_i f_\text{PEM2} \\
\label{eq:xphase}
x_i (t) & = \cos \left[ 2 \pi f_i (t - T_i) \right] \\
\label{eq:yphase}
y_i (t) & = \sin \left[ 2 \pi f_i (t - T_i) \right] \\
T_i & = \frac{(m_i T_\text{PEM1} + n_i T_\text{PEM2})}{(m_i f_\text{PEM1} + n_i f_\text{PEM2})}.
\end{align}
\end{subequations}

\noindent These in-phase and quadrature reference sinusoids, constructed for each harmonic $i$, are then multiplied by the mean-subtracted intensities $F_k(t)$ measured by the two PMTs $k$:
\begin{subequations}
\label{eq:xydc}
\begin{align}
DC_k & = \texttt{mean} \left[ F_k (t) \right] \\
\label{eq:xi}
X_{ik} & = \texttt{mean} \left \{ x_i (t) \left[ F_k (t) - \overline{F_k (t)} \right] \right \} / E_i \\
\label{eq:yi}
Y_{ik} & = \texttt{mean} \left \{ y_i (t) \left[ F_k (t) - \overline{F_k (t)} \right] \right \} / E_i.
\end{align}
\end{subequations}

\noindent $E_i$ are efficiency factors described below.

Since sinusoids are orthogonal, the means of the product of the reference sinusoids and the PMT signals (Equations \ref{eq:xi} and \ref{eq:yi}) pick out the intensity modulated at the harmonics of interest and reject intensity modulated at other frequencies. This is the key to the lock-in amplifier code, which is essentially a matched filter that reduces the bandwidth to $\sim 1$ Hz surrounding each of the 10 to 100 kHz harmonics. Time-varying signals not at the harmonics of interest are due to noise, generally assumed to be white in power spectral density, which is dramatically reduced by restricting the bandwidth around each harmonic. In contrast, an FFT does not make use of the reference frequencies, and its coarse frequency resolution enables significantly more noise to leak into the measurement.

After taking the means in Equations \ref{eq:xi} and \ref{eq:yi}, observables $X_i$, $Y_i$ (related to Stokes $Q$, $U$, and $V$), and $DC$ (related to Stokes $I$) are then calculated by division by the calculated modulation efficiency $E_i$ of each harmonic. Modulation efficiency indicates the polarization that would be measured were 100\% polarized light to be injected:
\begin{subequations}
\label{eq:stokesmodeff}
\begin{align}
\label{eq:stokesqmodeff}
E_\text{100} & = 2 J_2 (2.406 \text{ rad}) = 0.864 \\
\label{eq:stokesumodeff}
E_\text{10,90} & = 2 J_1 (1.451 \text{ rad}) J_1 (2.406 \text{ rad}) = 0.571 \\
E_\text{50} & = 2 J_0 (1.451 \text{ rad}) J_1 (2.406 \text{ rad}) = 0.559.
\end{align}
\end{subequations}

\noindent Here, $J_k (\beta_0)$ are Bessel functions of the $k$th kind that take the argument of the peak retardance setting $\beta_0$ of the PEMs. Since the PEMs are resonant devices, their retardance varies sinusoidally in time as $\beta (t) = \beta_0 \sin \omega_j (t - T_j)$ for PEM $j$. For PEM2, peak retardance is set as close as possible to the first zero of $J_0$, which is 2.4048 rad. However, the PEM controller does not have the accuracy to set the PEM to this value, so 2.406 rad is used and Stokes $Q$ modulation efficiency is 0.864 for the 100 kHz harmonic. While it could be increased to 0.971 for halfwave peak retardance $\beta_0 = \pi$, such as in the PlanetPol PEM \citep{Hough2006} and HIPPI-series polarimeter ferro-electric liquid crystal modulators \citep{Bailey2015}, such halfwave modulation exposes the system to unwanted systematic effects. That is, not only would Stokes $Q$ leak into Stokes $I$, but it would also magnify the instrumental conversion of $Q$ and $U$ between each other given a misalignment of the POLISH2 PEMs \citep{Hough2006, Wiktorowicz2008}. For PEM1, peak retardance of 1.451 rad is chosen to maximize modulation efficiency for the Stokes $u$ measurements.

Observables $X_i$, $Y_i$, and $DC$ are binned every 0.1 sec as an optimization to enable relatively high time resolution measurements but reduce computation time for data analysis. However, by recording raw data to disk, feasible given the low cost of data storage, subsequent re-reduction may be performed with any bin duration. For example, Crab Pulsar observations are typically reduced with bins of 1 ms duration or less. Another important benefit to recording raw data to disk and utilizing a lock-in amplifier in software, as opposed to using hardware lock-in amplifiers such as in PlanetPol and the POLISH prototype \citep{Wiktorowicz2008}, is that modulation at all important harmonics may be combined to increase the SNR of the polarization measurement. For example, not only does this enable simultaneous measurement of Stokes $Q$, $U$, and $V$ without rotating a waveplate or the telescope Cassegrain rotator, but it also enables the two independent measurements of Stokes $U$ at 10 and 90 kHz to be combined. This is crucial, because Equations \ref{eq:stokesqmodeff} and \ref{eq:stokesumodeff} show that modulation efficiency of a single Stokes $U$ harmonic in POLISH2 is only 66\% that of the Stokes $Q$ harmonic at 100 kHz. However, by combining both 10 and 90 kHz-derived Stokes $U$ estimates, which effectively multiplies Stokes $U$ modulation efficiency by a factor of $\sqrt{2}$, the Stokes $Q$ to $U$ modulation efficiency ratio increases to 94\%. Thus, software demodulation enables POLISH2 to simultaneously measure linear and circular polarization with relatively high modulation efficiency.

Each harmonic's phase $T_i$ (Equations \ref{eq:xphase} and \ref{eq:yphase}) is measured by injecting light through a linear or circular polarizer reference in the lab or on the telescope. These reference phases will differ even for harmonics that sample the same Stokes parameter (e.g., 10 and 90 kHz measurements of Stokes $u$). Thus, to combine polarization observables $X_i$ and $Y_i$ measured at 10 and 90 kHz, we rotate the $X_\text{10kHz}$ and $Y_\text{10kHz}$ phasor by the difference in reference phases measured at 10 and 90 kHz. Thus, the new reference phase measured at 10 kHz is identical to that measured at 90 kHz. At this point, $X_i$ and $Y_i$ of all harmonics for a given Stokes parameter may be combined. Practically speaking, this is only utilized for the Stokes $u$ 10 and 90 kHz harmonics, but other harmonics (Stokes $q$ 200 kHz; Stokes $u$ 110 and 190 kHz; and Stokes $v$ 30, 130, and 150 kHz) have been studied to extract maximum SNR from each measurement.

Fractional polarization is given by $q = Q/I$, $u = U/I$, and $v = V/I$. Therefore, harmonic-combined polarization observables $X_{QUV}$ and $Y_{QUV}$, indicating $X$ and $Y$ measured for each Stokes parameter $Q$, $U$, and $V$, are normalized by $DC$ in each 0.1 sec bin:

\begin{subequations}
\label{eq:quv}
\begin{align}
(Q,U,V) & \approx \sqrt{X_{QUV}^2 + Y_{QUV}^2} \\
I & = DC \\
\label{eq:stokesval}
(q,u,v) & \approx \sqrt{X_{QUV}^2 + Y_{QUV}^2} / DC \\
\label{eq:stokessign}
\Phi & = \texttt{atan2}(Y_{QUV},X_{QUV})
\end{align}
\end{subequations}

 \noindent Degree and orientation of linear polarization are given by $p \approx \sqrt{q^2 + u^2}$ and $\Theta = 1/2 \, \texttt{atan2(u,q)}$, respectively. Since degree of linear polarization $p$ is a positive definite quantity, the naive estimate $p = \sqrt{q^2 + u^2}$ is biased for low SNR data. The naive estimate $(Q,U,V) = \sqrt{X_{QUV}^2 + Y_{QUV}^2}$ is similarly biased. We therefore debias $p$, $Q$, $U$, and $V$ using the generalized MAS estimator \citep{Plaszczynski2014}, which is tabulated in Equations 7-9 of \cite{WiktorowiczNofi2015}. To calculate linear polarization orientation $\Theta$ in the correct quadrant, we utilize the commonly available function \texttt{atan2(u,q)} instead of $\arctan{(u/q)}$.

The sign of each Stokes parameter is related to the phase $\Phi$ of the $(X,Y)$ phasor (Equation \ref{eq:stokessign}) relative to that of a reference. That is, the Stokes parameter of interest is set to be positive for a phase difference between sample and reference of $| \Delta\Phi | < 90^\circ$ and negative for $| \Delta\Phi | > 90^\circ$. Here, $\Delta\Phi$ is wrapped to the range $\left[ -180^\circ, 180^\circ \right]$. Observations of lamp light through a linear and circular polarizer are used to measure the reference phase, and this is compared to the known Stokes parameter sign of strongly polarized stars on-sky. Since the reference phase is caused by the time delay between the PEM square wave reference signal for demodulation and corresponding light intensity variations, changes to hardware from run to run may cause a change in the reference phase. For example, one Stanford Research SR570 TIA failed on UT July 24, 2021, the first night of a Lick 3-m run, and the look-up table of phase references was updated when the new TIA was installed.

For observations of weakly polarized and/or faint targets, it is possible for uncertainty in AC phase $\Phi$ to cause the wrong sign of the Stokes parameter of interest to be applied, even if the absolute value of the Stokes parameter is accurate with respect to measured uncertainty. For this reason, subtraction of telescope polarization is performed on $X$ and $Y$, and conversion to Stokes parameters $q$, $u$, and $v$ is performed as late as possible in the data analysis process.

\subsection{Systematic Effects}
\subsubsection{Modulation Efficiency and Rotational Zero Point}

Linear polarization is a vector quantity, and both the magnitude and rotational zero point must be calibrated. The magnitude is calibrated by the modulation efficiency, which is the fractional polarization measured when $\sim 100\%$ polarized light is injected into the instrument. Measurement of linear polarization modulation efficiency may be performed by illuminating the POLISH2 entrance aperture with a light source and inserting a linear polarizer into the beam. This causes $\sim 100\%$ linearly polarized light to be injected. Additionally, we observe the \cite{Heiles2000} catalog of strongly polarized stars, which tend to be polarized at the percent level.

Since rotational zero point is tied to the precise mounting of the instrument at the beginning of each run, POLISH2 has mounting pins to locate its mounting plate on the telescope. Again, observation of strongly polarized standard stars, or guider images of astrometric fields, are used to determine the rotational zero point. For circular polarization, which is rotationally invariant and therefore requires no rotational zero point calibration, modulation efficiency and sign of polarization must be calibrated. A lamp and either a linear polarizer/quarter-wave Fresnel rhomb combination, or a circular polarizer, are used in series to inject $\sim100\%$ circularly polarized light into POLISH2.

\subsubsection{Telescope and Sky Polarization}
\label{sec:altaz}

Non-zero, telescope-induced polarization is a major systematic effect, and it is measurable even at straight Cassegrain focus. This well documented effect, due to incomplete cancellation of the polarization state of light generated by light rays across the telescope mirrors, tends to lie at the 100 ppm (0.01\%) level \citep{Hough2006, Wiktorowicz2008, Lucas2009, Wiktorowicz2009, Berdyugina2011, Bailey2015, WiktorowiczNofi2015, Wiktorowicz2015_189, Bott2016, Bailey2017, Bailey2020, Cotton2020, Marshall2020}. For equatorial or yoke-mounted telescopes, where the orientation of the telescope pupil on the sky is static with respect to pointing, telescope polarization may be calibrated by observation of nearly unpolarized stars. However, such telescopes tend to have long barrels, which can cause pointing-dependent mirror flexure. Thus, it is possible that telescope polarization may vary on rapid timescales and can be difficult to calibrate below the 10 ppm (0.001\%) level \citep{Wiktorowicz2015_189}.

\begin{deluxetable*}{cccccccccc}
\tabletypesize{\normalsize}
\tablecaption{POLISH2 Calibration Values}
\tablewidth{0pt}
\tablehead{
\colhead{Cal. Step} & \colhead{Item} & \colhead{100 kHz $q^\prime$} & \colhead{90 kHz $u^\prime$} & \colhead{10 kHz $u^\prime$} & \colhead{50 kHz $v^\prime$} & \colhead{$p^\prime$} & \colhead{$\Theta^\prime$ ($^\circ$)}}
\startdata
1		& Blue/Red PMT 1 		& 1.7432 & 0.7923 & 0.8370 & 0.9337 & $-$ & $-$ \\
$\cdots$	& Blue/Red PMT 2 		& 1.0000 & 1.0000 & 1.0118 & 1.0000 & $-$ & $-$ \\
\hline
2		& Lick 3-m 		& 0.7273 & 1.0000 & 1.0000 & 1.0000 & 0.5754 & $-$ \\
$\cdots$	& Lick 1-m 		& 0.7273 & 1.0000 & 1.0000 & 1.0000 & 0.5754 & 151.23 \\
$\cdots$	& Gemini Nov. 2016 	& 0.9935 & 1.0000 & 1.0000 & 1.0000 & 0.5754 & $-$ \\
$\cdots$	& Gemini Aug. 2018 	& 0.8324 & 1.0000 & 1.0000 & 1.0000 & 0.5754 & $-$
\label{tab:step12}
\enddata
\end{deluxetable*}

For alt-az telescopes, where the pupil rotates with parallactic angle, the linear polarization of either the target or telescope will vary sinusoidally with parallactic angle. By powering the Cassegrain de-rotator motor on, polarization of the target will be static in time while telescope polarization will rotate with parallactic angle. Conversely, powering the de-rotator off causes the opposite to occur. Thus, any target, even the science target itself, may in principle be used to calibrate telescope polarization \citep{Wiktorowicz2014, Millar-Blanchaer2020} provided the timescales of source and telescope polarization variations may be disentangled. This technique can be powerful for 8-10 m class alt-az telescopes for which maximizing science observations while maintaining proper calibration is critical. We demonstrate self calibration of science target polarization at Gemini North POLISH2 in section \ref{sec:gem}.

The last major systematic effect is background sky polarization. Though faint in the optical, the sky may reach linear polarization of $\sim 100\%$ when observing $90^\circ$ from the Moon. Since POLISH2's FOV is $19 \arcsec$ diameter at the Lick 3-m, significant sky polarization contamination may occur, especially for faint targets. Thus, POLISH2 performs an asymmetric sky nod cadence of target-sky-target ``triplets," where integrations are performed for 30 seconds at each nod position and 1/3 of each night is devoted to removal of background sky polarization. The sky nods are directed due north of the target field, and the throws at Gemini, the Lick 3-m, and the Lick 1-m are $15 \arcsec$, $30 \arcsec$, and $70 \arcsec$ to ensure no part of the target field overlaps with the sky field. While other polarimeters like the NOT 2.5-m TurPol subtract sky polarization more elegantly \citep{Piirola1973}, their use of choppers reduces throughput by 50\%. Thus, it is difficult to perform high accuracy polarimetry without sacrificing significant science throughput to enable sky polarization subtraction.

For POLISH2, we calculate mean linear and circular polarization of the sky during each 30 second sky integration, and sky polarization is interpolated in time using piecewise cubic Hermite interpolating polynomials at each time stamp of the on-target integrations. The slowest modulation frequency of POLISH2 is the 10 kHz modulation of Stokes $u$, and the fastest modulation is 100 kHz for Stokes $q$, so POLISH2 polarization measurements may be obtained with 0.01 to 0.1 millisecond temporal resolution without aliasing. As stated above, for computational efficiency, POLISH2 polarization measurements are typically binned every 0.1 second, though pulsar measurements are binned at millisecond or faster timescales. Thus, sky polarization is interpolated and subtracted from each 0.1 second on-target bin.

\section{POLISH2 Calibration}
\label{sec:labcal}
\subsection{Step 1: Calibration Across Modulation Frequency}
\label{sec:step1}

Simultaneous modulation of the beam by two photoelastic modulators causes independent polarization modulation at many beat-frequency harmonics. Instead of utilizing physical lock-in amplifiers that are sensitive to a single harmonic such as in PlanetPol \citep{Hough2006} and the POLISH prototype \citep{Wiktorowicz2008}, POLISH2's digitization of the signal from the PMTs improves SNR by measuring power at independent harmonics. That is, each POLISH2 modulation frequency provides an independent, simultaneous measurement of a given Stokes parameter: $f = 10$ and 90 kHz (Stokes $u$), 50 kHz (Stokes $v$), or 100 kHz (Stokes $q$). Each combination of Wollaston prism output beam and harmonic has a distinct modulation efficiency, which is due to both the intrinsic modulation efficiency $E_i$ from the physics of modulation (Equation \ref{eq:stokesmodeff}) and the frequency bandwidth of the detector, TIA, and DAQ system (Figure \ref{fig:fft}). POLISH polarization observables $X_{ik}$ and $Y_{ik}$, for harmonics $i$ and PMTs $k$, are divided by theoretical modulation efficiencies $E_i$ to calibrate for the former.

To calibrate for detector- and frequency-dependent modulation efficiency, we inject lamp light into POLISH2 through a linear polarizer (for Stokes $q$ and $u$), circular polarizer, or quarter wave Fresnel rhomb (for Stokes $v$), and we measure the resulting fractional polarization at each combination of detector and modulation frequency. This step is essentially like flatfielding an imaging detector, as it removes gain variations and ensures a uniform response to incident, polarized light. POLISH2 measurements are divided by these Step 1 calibration values, which are listed in Table \ref{tab:step12}. As these values are scaling factors to enable combination of polarization observables $X_{ik}$ and $Y_{ik}$ for harmonics $i$ and PMTs $k$, it is reasonable that some will be larger than unity. We find no evidence that these calibration values depend either on wavelength or on the pair of PMTs used (blue or red), but they do depend on the Wollaston prism output beam (1 or 2).

\subsection{Step 2: Calibration Across Stokes Parameters}
\label{sec:step2}

After Step 1 calibration is complete, the two PMTs deliver consistent Stokes parameters $q$, $u$, and $v$ at each modulation harmonic. Step 2 calibration ensures absolute calibration of those Stokes parameters both by accounting for differences in modulation efficiency between the two PEMs and by comparing linear polarization degree to the polarization catalogs of \cite{Heiles2000}. At the equatorially mounted Lick 3-m, observations are made at the same Cassegrain rotator angle from run to run. Since the analog measurement system of the rotator has an uncertainty of $0.1^\circ$ to $0.3^\circ$, instrumental Stokes $+q$ is typically consistent with Celestial $+q$ within this error. At the yoke mounted Lick 1-m, the Cassegrain rotator is not motorized. It may only be rotated manually, and it is nominally locked in position. The Lick 1-m is nominally clocked $\sim 30^\circ$ from Celestial $+q$, which is a rotational offset significant enough to nearly swap Stokes $q$ and $u$ from one telescope to the other. Thus, by comparing Lick 3-m and Lick 1-m POLISH2 observations of the same linearly polarized stars and at similar epochs, we may not only determine Stokes $q$, $u$, and $p$ modulation efficiencies, but we may also precisely determine the rotational offset between the two telescopes. That is, to accurately recover linear polarization $p_*$ and polarization position angle $\Theta_*$ of a star measured by POLISH2 at both the Lick 3-m and 1-m telescopes, only modulation efficiencies $q^\prime$ and $p^\prime$ as well as the Lick 1-m rotational offset $\Theta^\prime$ are required:
\begin{eqnarray}
q_{\rm{3m}} & = & p_* q^\prime p^\prime \cos{2\Theta_*} \\
u_{\rm{3m}} & = & p_* p^\prime \sin{2\Theta_*} \\
q_{\rm{1m}} & = & p_* q^\prime p^\prime \cos{2(\Theta_* - \Theta^\prime)} \\
u_{\rm{1m}} & = & p_* p^\prime \sin{2(\Theta_* - \Theta^\prime)} \\
p_* & = & 1/p^\prime \sqrt{(q_{\rm{3m}} / q^\prime)^2 + u_{\rm{3m}}^2} \\
\nonumber & = & 1/p^\prime \sqrt{(q_{\rm{1m}} / q^\prime)^2 + u_{\rm{1m}}^2} \\
\Theta^\prime & = & 1/2 \,\, {\texttt{atan2}}(u_{\rm{3m}},q_{\rm{3m}} / q^\prime) \\
\nonumber & - & 1/2 \,\, {\texttt{atan2}}(u_{\rm{1m}},q_{\rm{1m}} / q^\prime)
\end{eqnarray}

\begin{deluxetable*}{crcccccc}
\tabletypesize{\normalsize}
\tablecaption{Lick 3-m/1-m POLISH2 Comparison Stars}
\tablewidth{0pt}
\tablehead{
\colhead{Star} & \colhead{HD} & \colhead{R.A. (J2000)} & \colhead{Dec. (J2000)} & \colhead{$m_V$} & \colhead{Spec. Type} & \colhead{$p_{\rm{Heiles2000}}$ (\%)} & \colhead{$\Theta_{\rm{Heiles2000}}$ ($^\circ$)}}
\startdata
$\eta$ Per	& 17506	& 02 50 41.8	& +55 53 43.79	& 3.79	& K3-Ib-IIa	& 0.92(20)		& 118.0(6.2) \\
HD 41161	& 41161	& 06 05 52.5	& +48 14 57.43	& 6.76	& O8Vn		& 2.58(20)		& 169.0(2.2) \\
67 Oph	& 164353	& 18 00 38.7	& +02 55 53.60	& 3.93	& B5I		& 0.586(34)	& 69.3(1.7) \\
96 Her	& 164852	& 18 02 23.0	& +20 50 01.08	& 5.25	& B3IV		& 0.836(33)	& 171.8(1.1) \\
102 Her	& 166182	& 18 08 45.5	& +20 48 52.41	& 4.35	& B2IV		& 0.55(20)		& 173(10.3)
\label{tab:step2stars}
\enddata
\end{deluxetable*}

\begin{deluxetable*}{ccccccccc}
\tabletypesize{\normalsize}
\tablecaption{Absolute Polarization Comparison Stars}
\tablewidth{0pt}
\tablehead{
\colhead{Star} & \colhead{HD} & \colhead{R.A. (J2000)} & \colhead{Dec. (J2000)} & \colhead{$m_V$} & \colhead{Spec. Type} & \colhead{$p_{\rm{Catalog}}$ (\%)} & \colhead{$\Theta_{\rm{Catalog}}$ ($^\circ$)} & \colhead{Ref.}}
\startdata
$\eta$ Per    	& 17506	& 02 50 41.8	& +55 53 43.79	& 3.79	& K3-Ib-IIa	& 0.92(20)		& 118.0(6.2)	& 2 \\
HD 21291   	& 21291	& 03 29 04.1	& +59 56 25.21	& 4.22	& B9Ia		& 3.395(53)	& 115.30(40)	& 2 \\
$\epsilon$ Per	& 24760	& 03 57 51.2	& +40 00 36.78	& 2.89	& B1.5III		& 0.267(63)	& 13.9(6.7)	& 2 \\
$\zeta$ Tau   	& 37202	& 05 37 38.7	& +21 08 33.16	& 3.03	& B1IVe\_shell	& 1.54(23)		& 31.3(4.2)	& 2 \\
$\zeta$ Oph   	& 149757	& 16 37 09.5	& -10 34 01.53	& 2.56	& O9.2IVnn	& 1.295(23)	& 126.10(50)	& 2 \\
HD 154445  	& 154445	& 17 05 32.3	& -00 53 31.44	& 5.61	& B1V		& 3.780(62)	& 88.79(47)	& 1 \\
HD 157999  	& 157999	& 17 26 30.9	& +04 08 25.29	& 4.31	& K2II		& 1.010(35)	& 85.9(1.0)	& 2 \\
HD 161056  	& 161056	& 17 43 47.0	& -07 04 46.59	& 6.32	& B3II/III		& 4.030(25)	& 66.93(18)	& 1 \\
67 Oph     		& 164353	& 18 00 38.7	& +02 55 53.60	& 3.93	& B5I		& 0.586(34)	& 69.3(1.7)	& 2 \\
96 Her     		& 164852	& 18 02 23.0	& +20 50 01.08	& 5.25	& B3IV		& 0.7898(80)	& 171.22(29)	& 3 \\
102 Her    		& 166182	& 18 08 45.5	& +20 48 52.41	& 4.35	& B2IV		& 0.3799(30)	& 173.00(23)	& 3 \\
HD 176155  	& 176155	& 18 58 14.7	& +17 21 39.30	& 5.38	& F6Ib		& 0.6200(60)	& 32.80(30)	& 2 \\
HD 176818  	& 176818	& 19 01 17.8	& +21 30 49.87	& 7.05	& B1V		& 1.0433(86)	& 31.11(24)	& 3 \\
HD 183143  	& 183143	& 19 27 26.6	& +18 17 45.19	& 6.86	& B6Ia		& 5.886(64)	& 178.80(30)	& 2 \\
$\kappa$ Aql  	& 184915 	& 19 36 53.4	& -07 01 38.92	& 4.96	& B0.5IIIn		& 1.35(33)		& 171.9(7.0)	& 2 \\
10 Sge     		& 188727	& 19 56 01.3	& +16 38 05.24	& 5.36	& F7Ib		& 0.6890(90)	& 15.50(40)	& 2 \\
HD 193237  	& 193237	& 20 17 47.2	& +38 01 58.55	& 4.82	& B1-2Ia-0ep	& 1.11(20)		& 34.0(5.1)	& 2 \\
55 Cyg     		& 198478	& 20 48 56.3	& +46 06 50.88	& 4.86	& B4Ia		& 2.877(90)	& 2.50(90)		& 2 \\
$\sigma$ Cyg  	& 202850	& 21 17 25.0	& +39 23 40.85	& 4.24	& A0Ia		& 0.569(49)	& 15.0(2.5)	& 2 \\
HD 204827  	& 204827	& 21 28 57.8	& +58 44 23.22	& 7.94	& O9.5IV		& 5.322(14)	& 58.730(80)	& 1 \\
HD 207673  	& 207673	& 21 49 40.1	& +41 08 55.63	& 6.48	& A2Ib		& 0.329(22)	& 140.4(1.9)	& 2 \\
\label{tab:step2pstars}
\enddata
\tablerefs{1. \cite{Schmidt1992}, 2. \cite{Heiles2000}, 3. \cite{Weitenbeck2004}}
\end{deluxetable*}

The five stars in Table \ref{tab:step2stars} were observed at the Lick 3-m from February 4, 2012 UT to October 14, 2013 UT and at the Lick 1-m from March 23, 2012 UT to May 30, 2014 UT. For the tables in this paper, values in parenthesis denote uncertainty in the last two digits. To determine $q^\prime$, we divide $q_{\rm{3m}}$ and $q_{\rm{1m}}$, Stokes $q$ measured at both telescopes, by a grid of $q^\prime$ values. Degree of linear polarization $p$ and polarization position angle $\Theta$ are re-calculated for each star at each grid point, and the difference $\Delta \Theta$ is calculated for each star between both telescopes. The true value $q^\prime = 0.7273$ (Table \ref{tab:step12}) minimizes the square root of the weighted variance of $\Delta \Theta$ across the five-star sample. Once $q^\prime$ is determined, it is straightforward to determine the rotational offset of the Lick 1-m with respect to the Lick 3-m to be $\Theta^\prime = 151.23^\circ = -28.77^\circ$ (Table \ref{tab:step12}). At this point, the entire suite of polarized stars observed is compared to catalogs \citep{Schmidt1992, Heiles2000, Weitenbeck2004} to calibrate POLISH2 for absolute polarization (Table \ref{tab:step2pstars}). This leads to the determination that $p^\prime = p_{\rm{POLISH2}} / p_{\rm{Catalogs}} = 0.5754$ (Table \ref{tab:step12}). This value is nearly equal to 0.5, which accounts for an unnecessary factor of 2 located in the POLISH2 data reduction pipeline.

All POLISH2 Stokes parameters and uncertainties are therefore divided by the Step 1 and 2 calibration factors $q^\prime$, $u^\prime$, $v^\prime$, and $p^\prime$ in Table \ref{tab:step12}. Lick 1-m POLISH2 data are then rotated by $\Theta^\prime$, and Gemini North POLISH2 data are discussed in a later section. To test calibration accuracy, POLISH2 Stokes parameters from the stars in Table \ref{tab:step2pstars} are again compared to catalogs after the above calibration factors are applied. Table \ref{tab:pcalerr} shows that the weighted mean ratio $p^\prime = p_{\rm{POLISH2}} / p_{\rm{Catalogs}} = 1.010 \pm 0.020$ ($1\sigma$) for the ensemble of polarized stars after calibration. Thus, POLISH2's absolute accuracy is 1\% with a $1\sigma$ range of 2\% on that estimate. The limiting factor in the accuracy of POLISH2 linear polarization calibration is the uncertainty on individual stellar polarization in catalogs.

\begin{deluxetable}{cccc}
\tabletypesize{\normalsize}
\tablecaption{Calibration Factor Residuals}
\tablewidth{0pt}
\tablehead{
\colhead{Band} 		& \colhead{Residual $p^\prime$}}
\startdata
$B$					& 1.012(17) \\
$V$					& 0.978(32) \\
Clear				& 0.978(44) \\
\hline
Overall				& 1.010(20)
\label{tab:pcalerr}
\enddata
\end{deluxetable}

\begin{figure}
\centering
\includegraphics[scale=0.34]{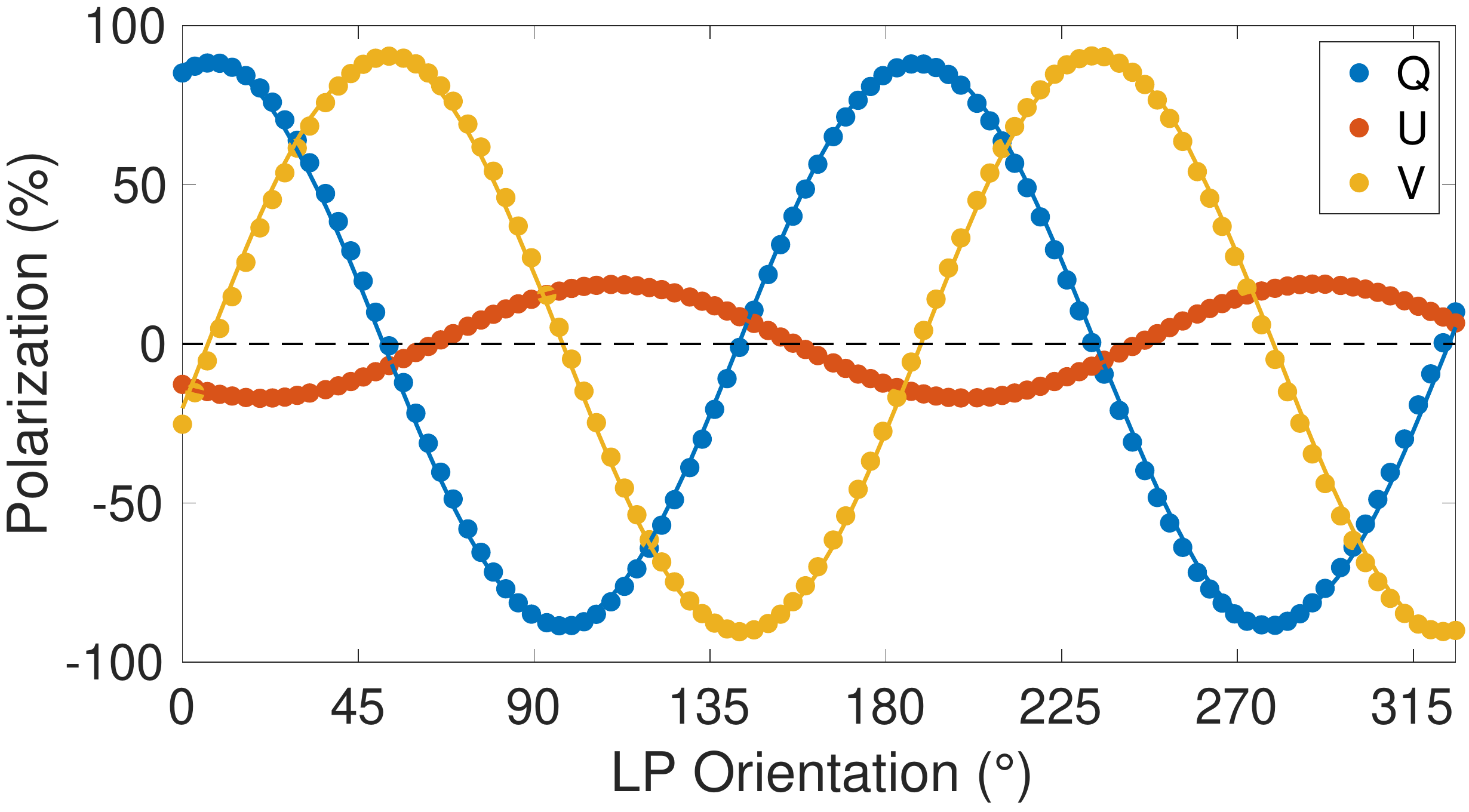}
\caption{Simultaneous measurement of linear (Stokes $q$ and $u$) and circular polarization (Stokes $v$) obtained in the lab with POLISH2 by injecting light through a rotating linear polarizer (LP) and quarter wave Fresnel rhomb in series. This polarization modulation is measured in real time during the $\sim1.5$ minute long polarizer rotation. Data points are binned every 0.9 sec, and the solid curves represent fits to Equations \ref{eq:fresfit1} though \ref{eq:fresfit2}.}
\label{fig:fresmod}
\end{figure}

To calibrate for circular polarization, both a right-circular and left-circular polarizer are inserted in the beam from a light source in the lab, and a high modulation efficiency of 93\% is measured (Table \ref{tab:step12}). In addition, POLISH2 accurately recovers the sign change between both circular polarizers. To further demonstrate POLISH2's accuracy in measuring circular polarization, a quarter-wave Fresnel rhomb is inserted between a linear polarizer and the POLISH2 entrance aperture in the lab. The linear polarizer is installed in a Newport AG-PR100 motorized rotation stage and the Fresnel rhomb is installed in a manual rotation stage. Data are obtained while the polarizer rotates to measure sinusoidal modulation of Stokes $q$, $u$, and $v$ simultaneously (Figure \ref{fig:fresmod}). After each $\sim 360^\circ$ rotation of the linear polarizer, the Fresnel rhomb itself is incrementally rotated by hand about the optical axis, and a new integration is obtained while the linear polarizer rotates. This injects the following Stokes parameters into POLISH2 for linear polarizer and Fresnel rhomb orientations $\Theta$ and $\phi$, respectively:
\begin{subequations}
\begin{align}
\label{eq:fresfit1}
q &= 1/2 \left[\cos{(2 \Theta + 4 \phi)} + \cos{2 \Theta} \right] \\
u &= 1/2 \left[\sin{(2 \Theta + 4 \phi)} - \sin{2 \Theta} \right] \\
v &= -\sin{(2 \Theta + 2 \phi)}
\label{eq:fresfit2}
\end{align}
\end{subequations}

\noindent Peak $q$ and $u$ occur at $\Theta = -\phi$, and peak $v$ occurs at $\Theta = \pi/4 - \phi$. The maximum values of the Stokes parameters at these linear polarizer orientations are given by the following:
\begin{subequations}
\begin{align}
\max{(q)} &= |\cos{2 \phi}| \\
\max{(u)} &= |\sin{2 \phi}| \\
\max{(v)} &= 100\%
\end{align}
\end{subequations}

\noindent Since this setup injects POLISH2 with a source of assumed 100\% circular polarization, modulation efficiency for circular polarization (Stokes $v$) may be obtained. Additionally, it is straightforward to derive that the orientation of the Fresnel rhomb $\phi$ may be recovered from the following ratio:
\begin{eqnarray}
 \label{eq:frescalc}
 \max{(u)} / \max{(q)} = |\tan{2 \phi}|
\end{eqnarray}

 Observation of this effect would conclusively demonstrate the ability of POLISH2 to accurately measure circular polarization. Since the motorized rotation stage does not have direct encoder feedback, absolute orientation of the linear polarizer is not known a priori. Thus, these measurements are necessary to derive absolute orientation of the linear polarizer $\Theta$ and Fresnel rhomb $\phi$.

Table \ref{tab:fres} shows the results of measurements obtained at four rotational orientations of the Fresnel rhomb, where the linear polarizer was rotated through  $\sim 360^\circ$ while data were being recorded at each orientation of the Fresnel rhomb. The first column in this table shows the value of each Fresnel rhomb orientation, which was recorded in micrometer units of the Fresnel rhomb rotation stage. The second column shows the ratio of the maximum Stokes $u$ value to the maximum Stokes $q$ value during rotation of the linear polarizer. The third column lists the true Fresnel rhomb orientations in degrees calculated from Equation \ref{eq:frescalc}. Finally, the last column shows Stokes $v$ modulation efficiency after Step 1 and 2 calibration values in Table \ref{tab:step12} are applied.

\begin{deluxetable}{cccc}
\tabletypesize{\normalsize}
\tablecaption{Fresnel Rhomb Results}
\tablewidth{0pt}
\tablehead{
\colhead{$\phi$ Meas.} & \colhead{$\max{(u)} / \max{(q)}$} & \colhead{$\phi$ Calc.} & \colhead{$v$ Mod. Eff.} \\
(Units) & & ($^\circ$) & ($\%$)}
\startdata
       $-1.0$ & 0.3936 & 10.741 & 98.94 \\
       $-0.5$ & 0.2887 & 8.051 & 100.82 \\
       0.0 & 0.1850 & 5.242 & 100.04 \\
       0.5 & 0.0883 & 2.522 & 100.09
\label{tab:fres}
\enddata
\end{deluxetable}

\begin{figure}
\centering
\includegraphics[scale=0.34]{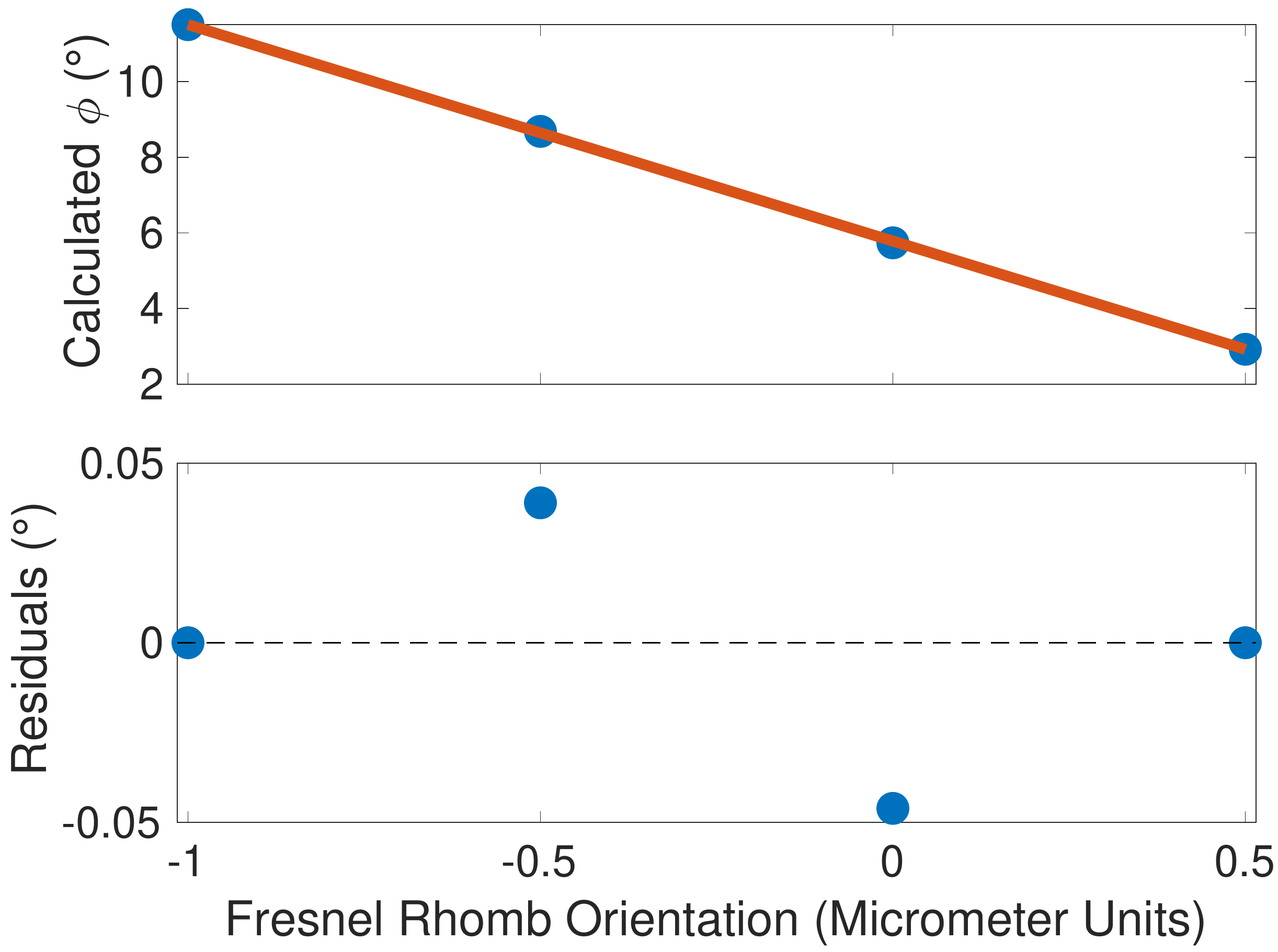}
\caption{\textit{Top:} Laboratory Fresnel rhomb orientation calculated from the relative amplitude of Stokes $q$ and $u$ measurements via Equation \ref{eq:frescalc}. The best fit line, $\Theta = (-5.726^\circ \pm 0.050^\circ) \times \Theta_\text{micrometer} + (5.783^\circ \pm 0.028^\circ)$, is shown in red. \textit{Bottom:} Residuals to the linear fit are shown, and the standard deviation of the residuals is only $0.035^\circ$.}
\label{fig:frescalc}
\end{figure}

Figure \ref{fig:frescalc} shows the remarkable linear correlation between measured Fresnel rhomb orientation in micrometer units and the calculated orientation in degrees. This correlation is given by 
\begin{eqnarray}
 \label{eq:frescal}
 \Theta & = & (-5.726^\circ \pm 0.050^\circ) \times \Theta_\text{micrometer} \\
\nonumber & + & (5.783^\circ \pm 0.028^\circ),
\end{eqnarray}

\noindent and the standard deviation of the residuals is only $0.035^\circ$ over the $8.589^\circ$ range of the experiment. Thus, measurements of circular polarizers and a quarter wave Fresnel rhomb illustrate that POLISH2's measurements of circular polarization are of high accuracy.

\subsection{Step 3: Calibration of Instrumental Crosstalk}
\label{sec:step3}

Unwanted instrumental conversion of linear to circular polarization (``crosstalk") significantly hampers the accuracy of circular polarimetry. In a waveplate polarimeter with reimaging optics upstream of the waveplate, crosstalk may be caused by phase retardance introduced by off-axis reflection of light by coated optics. It can also be caused by static stress birefringence frozen into lenses by the annealing process. However, it will certainly be due in part to non-ideal retardance variations with wavelength that are present even in superachromatic waveplates. That is, superachromatic, visible light, half waveplates typically have retardance of $0.50 \pm 0.01$ waves, which will cause wavelength-dependent crosstalk with efficiency between $\pm 12.5\%$. Thus, observations of a $\sim1\%$ linearly polarized star will include spurious circular polarization, caused by the waveplate, that vary with wavelength between $\pm 0.1\%$.

\begin{figure}
\centering
\includegraphics[scale=0.34]{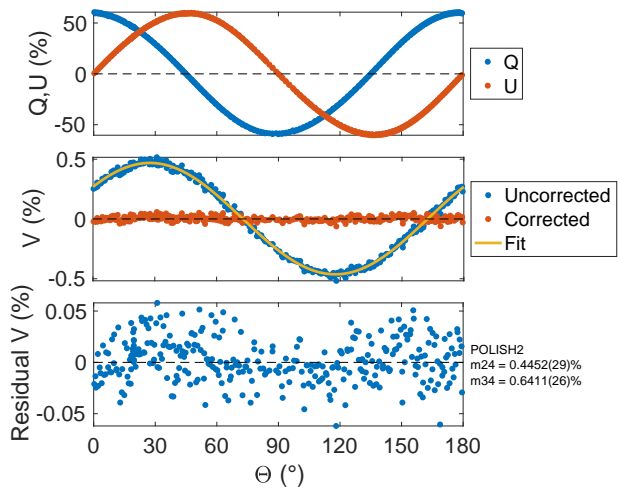}
\caption{Simultaneous linear and circular polarimetry of the daytime sky above the Lick 3-m measured with POLISH2 while the Cassegrain rotator was continuously rotated. \textit{Top:} Linear polarization (Stokes $q$ and $u$) varies sinusoidally as a function of polarization orientation $\Theta = \texttt{atan2(u,q)}$. \textit{Middle:} Circular polarization (Stokes $v$) varies sinusoidally as a function of $\Theta$, which must be due to instrumental crosstalk (blue points) since intrinsic circular polarization is invariant under rotation. Utilization of the POLISH2 Mueller matrix (yellow curve, Equation \ref{eq:crosstalk}), yields corrected circular polarization (red points). Note the factor of $\sim 100$ scale difference between the top and middle panels. \textit{Bottom:} Residual circular polarization has a standard deviation of 0.02\%.}
\label{fig:crosstalk}
\end{figure}

Even though POLISH2 utilizes photoelastic modulators instead of waveplates, crosstalk of unknown origin still clearly exists in the system. Figure \ref{fig:crosstalk} shows simultaneous linear and circular polarimetry obtained on the sunlit, late afternoon sky above the Lick 3-m telescope while the Cassegrain rotator was continuously rotating. Circular polarization (Stokes $v$) of $\sim 0.5\%$ amplitude, but varying sinusoidally with linear polarization orientation $\Theta$, is superimposed on the $\sim 50\%$ linearly polarized sky (Stokes $q$ and $u$). While the sky has been shown to harbor intrinsic circular polarization of order $v \sim 0.1\%$ \citep{Angel1972}, such intrinsic circular polarization would be invariant under rotation of the Cassegrain rotator. Indeed, any circular polarization generated by the Lick 3-m mirrors would also be invariant under rotation of POLISH2. Thus, variations in measured circular polarization in Figure \ref{fig:crosstalk} must be intrinsic to the POLISH2 instrument.

Mathematically, the Mueller matrix of a generic optic (such as the entire POLISH2 system) is given by the following:
\begin{eqnarray}
\label{eq:nonidealpol}
M_\text{generic} = \begin{bmatrix} m_{11} & m_{12} & m_{13} & m_{14} \\ m_{21} & m_{22} & m_{23} & m_{24} \\ m_{31} & m_{32} & m_{33} & m_{34} \\ m_{41} & m_{42} & m_{43} & m_{44} \end{bmatrix}.
\end{eqnarray}

\noindent Measured circular polarization of the sky in Figure \ref{fig:crosstalk} is consistent with the following equation:
\begin{eqnarray}
 \label{eq:crosstalk}
v_\text{meas} & = & v_0 + q_0 (m_{42} \cos 2 \Theta + m_{43} \sin 2 \Theta) \\
\nonumber & + & u_0 (m_{43} \cos 2 \Theta - m_{42} \sin 2 \Theta).
\end{eqnarray}

\noindent Here, measured circular polarization $v_\text{meas}$ is dependent on intrinsic linear and circular polarization of the sky ($q_0$, $u_0$, and $v_0$) as well as the POLISH2 Mueller matrix elements $m_{42}$ (conversion of Stokes $q$ to $v$) and $m_{43}$ (conversion of Stokes $u$ to $v$) from Equation \ref{eq:nonidealpol}.  Fitting Equation \ref{eq:crosstalk} to the data in Figure \ref{fig:crosstalk}, we measure intrinsic circular polarization of the sky to be $v_0 = 28 \pm 12$ ppm ($0.0028 \pm 0.0012\%$).

 Thus, the highly correlated, sinusoidal variation of Stokes $v$ as a function of linear polarization orientation $\Theta$ that is measured by POLISH2 requires conversion of linear polarization to circular polarization with $\sim 1\%$ efficiency. Lab measurements with POLISH2 and a rotating linear polarizer require similar values of $m_{42}$ and $m_{43}$ as in Figure \ref{fig:crosstalk}, which identifies the POLISH2 PEMs as the source of crosstalk. The polarimeter of \cite{Kemp1972} utilizes a single PEM, and measured crosstalk at the level of $0.3\%$ is hypothesized to be caused by residual strain-induced birefringence. This is similar to the value measured for POLISH2. However, POLISH2's crosstalk with $\sim 1\%$ efficiency is an order of magnitude less severe that those of superachromatic half waveplates.
 
\begin{figure}
\centering
\includegraphics[scale=0.34]{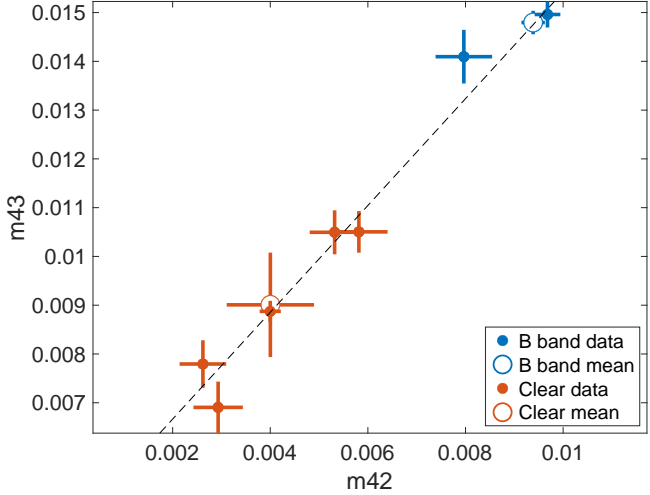}
\caption{POLISH2 Mueller matrix elements indicating crosstalk measured during continuous rotation of the Lick 3-m Cassegrain rotator under the daytime sky. Unwanted instrumental conversion of Stokes $u$ to $v$, indicated by $m_{43}$, is plotted against Stokes $q$ to $v$ conversion $m_{42}$. A linear fit to the data indicates a striking, wavelength-dependent trend (dashed line).}
\label{fig:mueller}
\end{figure}

After seven measurements of the POLISH2 Mueller matrix at the Lick 3-m via the daytime sky between July and September 2021, we find that $m_{42}$ and $m_{43}$ vary with wavelength by roughly a factor of two between $B$ and unfiltered bandpasses (Figure \ref{fig:mueller} and Table \ref{tab:mueller}). We also find these Mueller matrix elements to be similar from run to run, so the mean values in Table \ref{tab:mueller} are used to correct for POLISH2 crosstalk in all POLISH2 data. To do so, we take advantage of the following equation describing how incident polarization $S_\text{incident}$ is altered by the POLISH2 Mueller matrix $M_\text{POLISH2}$ prior to measurement:
\begin{subequations}
\begin{align}
\tiny \begin{bmatrix} 1 \\ q_\text{meas} \\ u_\text{meas} \\ v_\text{meas} \end{bmatrix} & = M_\text{POLISH2} \times \tiny \begin{bmatrix} 1 \\ q_\text{incident} \\ u_\text{incident} \\ v_\text{incident} \end{bmatrix} & \\
S_\text{meas} & = M_\text{POLISH2} \times S_\text{incident} & \\
S_\text{incident} & = M^{-1}_\text{POLISH2} \times S_\text{meas} & \\
 \label{eq:mueller1}
M_\text{POLISH2} & = \begin{bmatrix} 1 & 0 & 0 & 0 \\ 0 & 1 & 0 & +0.0094 \\ 0 & 0 & 1 & +0.0148 \\ 0 & +0.0094 & +0.0148 & 1 \end{bmatrix} & \\
 \label{eq:mueller}
S_\text{incident} & = \tiny \begin{bmatrix} 1 & 0 & 0 & 0 \\ 0 & 1 & 0 & -0.0094 \\ 0 & 0 & 1 & -0.0148 \\ 0 & -0.0094 & -0.0148 & 1 \end{bmatrix} \begin{bmatrix} 1 \\ q_\text{meas} \\ u_\text{meas} \\ v_\text{meas} \end{bmatrix} &
\end{align}
\end{subequations}

\begin{deluxetable*}{cccccc}
\tabletypesize{\normalsize}
\tablecaption{POLISH2 Mueller Matrix Measurements}
\tablewidth{0pt}
\tablehead{
\colhead{UT Date} & \colhead{Band}	& \colhead{$m_{42}$} & \colhead{$m_{43}$} & \colhead{$p_0$ (\%)} & \colhead{$v_0$ (ppm)}}
\startdata
2021 Jul 26 00:35	& $B$	& 0.00970(26)	& 0.01496(27)	& 17.81(14)	& 76(36) \\
2021 Aug 14 02:57	& $B$	& 0.00799(59)	& 0.01399(56)	& 18.803(37)	& $-$11(83) \\
\hline
Mean	 & $B$	& 0.00940(24)	& 0.01477(24)	& $-$	& $-$ \\
\hline
2021 Jul 26 00:45	& Clear	& 0.00529(53)	& 0.01033(47)	& 4.040(48)	& 23(15) \\
2021 Jul 26 00:49	& Clear	& 0.00588(59)	& 0.01044(43)	& 3.431(46)	& 27(12) \\
2021 Aug 14 03:19	& Clear	& 0.00400(22)	& 0.00888(21)	& 22.845(59)	& $-101(36)$ \\
2021 Sep 14 00:16	& Clear	& 0.00293(51)	& 0.00690(53)	& 42.804(39)	& 80(170) \\
2021 Sep 14 00:19	& Clear	& 0.00262(48)	& 0.00779(49)	& 43.527(23)	& $-50(170)$ \\
\hline
Mean	 & Clear	& 0.00394(90)	& 0.0089(10)	& $-$	& $-$
\label{tab:mueller}
\enddata
\end{deluxetable*}

\noindent Here, the POLISH2 Mueller matrix for $B$ band is tabulated, and we assume $m_{24} = m_{42}$ and $m_{34}$ = $m_{43}$ for symmetry. Elements $m_{24}$ and $m_{34}$ represent circular to linear polarization conversion. Accurate measurement of these Mueller matrix elements would require a calibrated source of significant linear and circular polarization simultaneously, which is beyond the scope of this paper. Figure \ref{fig:crosstalk} suggests that no significant conversion between Stokes $(q,u)$ and $(u,q)$ exists, so we assume $m_{23} \sim m_{32} \sim 0$ in $M_\text{POLISH2}$ (Equation \ref{eq:mueller1}). A striking, wavelength-dependent correlation is present between $m_{42}$ and $m_{43}$ in Figure \ref{fig:mueller}, which suggests that the POLISH2 Mueller matrix may be predicted for any bandpass.

Utilizing Equation \ref{eq:mueller}, we demonstrate that spurious circular polarization introduced by POLISH2 may be corrected with high accuracy (Figure \ref{fig:crosstalk}). Indeed, the standard deviation of residual circular polarization, measured on-sky during rotation of the Lick 3-m Cassegrain rotator, is 0.02\%. Thus, for incident linear polarization of $\sim 50\%$, accurate circular polarization may be obtained with crosstalk mitigated to the level of one part in up to 2,500. This suggests that objects with a ratio of circular to linear polarization measured by POLISH2 to be greater than 1/2,500 harbor intrinsic circular polarization. Indeed, for objects with 1\% linear polarization, the standard deviation of residual, crosstalk-induced circular polarization will only be 4 ppm.

\subsection{Non-Ideal Polarization from Lab Optics}
\subsubsection{Crossed Linear Polarizers}

After demonstrating control of instrumental systematic effects in linear and circular polarization, we demonstrate the ability to accurately measure systematic effects intrinsic to lab optics themselves. By injecting light through two thin-film linear polarizers sequentially in the lab, where the upstream polarizer is inserted in a motorized rotation stage while the downstream one is held static, we measure crosstalk intrinsic to the linear polarizers. For an ideal polarizer, $m_{11} = m_{12} = m_{21} = m_{22} = 0.5$ in Equation \ref{eq:nonidealpol}, and all other Mueller matrix elements are zero. It is straightforward to derive that regardless of the polarization state of incident light, the fractional polarization output of an ideal linear polarizer is always 100\% and aligned parallel to the orientation of the polarizer ($q = Q/I = 0.5 / 0.5 = 100\%$). No light will pass through a pair of ideal linear polarizers when perfectly crossed, and therefore this output polarization state is meaningless. In this orientation, all Mueller matrix elements of the system containing the pair of crossed polarizers are zero.

\begin{figure}
\centering
\includegraphics[width=0.47\textwidth]{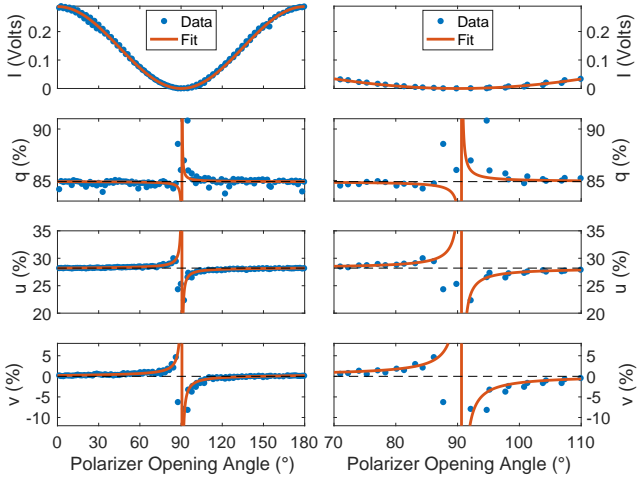}
\caption{POLISH2 lab measurements of light injected through two sequential, linear polarizers as the upstream polarizer rotates continuously. For an opening angle of $90^\circ$, the two polarizers are crossed. Two ideal polarizers would show constant Stokes $q$, $u$, and $v = 0$ as the opening angle varies, so observed discontinuities in Stokes parameters require crosstalk intrinsic to the polarizers. The right panels zoom in on the discontinuity observed when the polarizers are perfectly crossed, and the vertical scale is maintained between left and right panels.}
\label{fig:crossedpol}
\end{figure}

Figure \ref{fig:crossedpol} shows POLISH2 $B$ band measurements obtained as the upstream polarizer rotates from a parallel orientation to a crossed one with respect to the static, downstream polarizer. These data have been corrected for POLISH2 systematic effects via Step 1, 2, and 3 calibrations (Sections 
\ref{sec:step1} through \ref{sec:step3}). Clearly, discontinuities in $q$, $u$, and $v$ occur in the $\pm 5^\circ$ surrounding the crossed orientation, which cannot occur in ideal polarizers (save perhaps for the instant that zero light passes through the system). These trends therefore cannot be caused by uncorrected systematic effects in POLISH2. This is because the downstream linear polarizer is held static, and were it an ideal polarizer, it would output constant Stokes $q$ and $u$ (with $v = 0$) regardless of the orientation of the upstream, rotating linear polarizer. Thus, we conclusively demonstrate various manners of crosstalk intrinsic to all thin-film linear polarizers themselves by the following methodology.

In the frame of the POLISH2 Wollaston prism, which sets the orientation zero point of POLISH2, we model the system containing the following: a light source assumed to be unpolarized, a pair of non-ideal but identical linear polarizers (where the downstream polarizer is held static at an orientation $\phi$ and the upstream polarizer rotates with an opening angle $\theta$ with respect to the downstream polarizer), and POLISH2. This system is described by the following:
\begin{widetext}
\begin{subequations}
\begin{align}
\label{eq:crosspolM}
S_\text{meas} & = M_\text{POLISH2} \times T(-\phi) \times M_\text{LP} \times T(-\theta) \times M_\text{LP} \times T(\theta) \times T(\phi) \times S_\text{incident} & \\
\label{eq:crossedpol}
I & = (1 + \cos 2 \theta) / 4 & \\
q = Q/I & = \cos 2 \phi - (2 M_{33} \sin 2 \theta \sin 2 \phi) / (1 + \cos 2 \theta) & \\
u = U/I & = \sin 2 \phi + (2 M_{33} \sin 2 \theta \cos 2 \phi) / (1 + \cos 2 \theta) & \\
v = V/I & = 2 M_{43} \sin 2 \theta / (1 + \cos 2 \theta). &
\end{align}
\end{subequations}
\end{widetext}

Here, Mueller matrix elements $M_{ij}$ of the linear polarizers are capitalized to distinguish them from POLISH2 Mueller matrix elements $m_{ij}$, and rotation matrices are given by $T$. The simplest model to acceptably reproduce observed polarization in Figure \ref{fig:crossedpol} requires only Mueller matrix elements $M_{33}$ and $M_{43}$ intrinsic to the linear polarizers themselves. All other Mueller matrix elements, both from the linear polarizers and from POLISH2 residuals after the above calibrations are performed, do not appear at a level significant enough to justify their inclusion in the model. To determine the Mueller matrix of the non-ideal linear polarizers, and to identify the major matrix elements necessary to describe measured trends, we perform the following procedure. As stated above, the motorized linear polarizer rotation stage does not have direct encoder feedback, so the opening angle $\theta$ is mapped to observation time by fitting for intensity $I$ via Equation \ref{eq:crossedpol}. Here, $\theta = 90^\circ$ (perfectly crossed polarizers) occurs when the minimum light level is measured (Figure \ref{fig:crossedpol}, top panels). Assuming stray light is mitigated properly, the ratio of maximum to minimum light level indicates the linear polarizer extinction ratio in $B$ band. This value is about 670, which is reasonable among low-cost thin film polarizers. Thus, such polarizers transmit $1/670 = 0.15\%$ of incident Stokes $-Q$ when oriented in the Stokes $+Q$ direction.

The presence of Stokes $u$ even for parallel linear polarizers ($\theta = 0^\circ$ and $180^\circ$) requires the downstream polarizer to be canted by $8.1925949 \pm 0.0000013^\circ$ with respect to the POLISH2 Wollaston prism. Then, the variations of Stokes $q$ and $u$ surrounding the $\theta = 90^\circ$ discontinuity require $M_{33} = 0.0005836 \pm 0.0000033$. Finally, the presence of Stokes $v$ surrounding the $\theta = 90^\circ$ discontinuity requires linear to circular conversion (specifically, Stokes $u$ to $v$) intrinsic to the polarizers themselves with $M_{43} = 0.00145693 \pm 0.00000086$. Note that this value is an order of magnitude smaller than that intrinsic to POLISH2 (Figure \ref{fig:mueller}), yet it may be measured with high accuracy. Thus, only the Mueller matrix elements $M_{33}$ and $M_{43}$, intrinsic to the polarizers, are required to simultaneously explain observed Stokes $q$, $u$, and $v$. This demonstrates that our calibration approach in Section \ref{sec:step3} accurately captures POLISH2's intrinsic crosstalk.

After Equation \ref{eq:nonidealpol}, the following represents the measured Mueller matrix of generic thin-film linear polarizers, where $M_{33} \sim 0.0006$ and $M_{43} \sim 0.0015 $ are commonly assumed to be zero:

\begin{eqnarray}
\label{eq:nonidealpol2}
M_\text{LP} = \begin{bmatrix} 0.5 & 0.5 & 0 & 0 \\ 0.5 & 0.5 & 0 & 0 \\ 0 & 0 & 0.0006 & 0 \\ 0 & 0 & 0.0015 & 0 \end{bmatrix}.
\end{eqnarray}

\noindent This matrix may be visualized in the following manner. Unpolarized light with $q = u = v = 0$ propagates through the system from the right side of Equation \ref{eq:crosspolM} to the left. Then, the rotation matrices $T(\theta) \times T(\phi)$ and the ideal portion of the first linear polarizer, $M_{11} = M_{12} = M_{21} = M_{22} = 0.5$, generate non-zero $q$ and $u$ but retain $v = 0$. The second linear polarizer, if ideal, would remove $u$ and output $q$ into POLISH2, which is rotated by $\phi$ with respect to the second polarizer. However, the second non-ideal polarizer transmits $M_{33} \sim 0.06\%$ of incident $u$, and it converts $M_{43} \sim 0.15\%$ of incident $u$ into $v$. Thus, as the first polarizer presents the second polarizer with time-varying $q$ and $u$, the second polarizer outputs $q$, $u$, and $v$ into POLISH2.

\subsubsection{Integrating Spheres and Cavity Blackbodies}

\begin{deluxetable*}{cccccccc}
\tabletypesize{\small}
\tablecaption{Lab Source Polarization}
\tablewidth{0pt}
\tablehead{
\colhead{Source} & \colhead{UT Date} & \colhead{Band} & \colhead{$q$ (ppm)} & \colhead{$u$ (ppm)} & \colhead{$v$ (ppm)} & \colhead{$p$ (ppm)} & \colhead{$\Theta$ $(^\circ)$}}
\startdata
IR Industries BB\tablenotemark{a} & 2016 Apr 21 	 & $B$	& 87(31)       	 & $-$70(30)     	 & $-$174(63)   	 & 108(31)    	 & $-$19.5(8.1) \\
IR Industries BB & 2016 Apr 21 	 & $V$	& 25.8(8.7)    	 & $-$101.0(8.9) 	 & $-$163(22)   	 & 103.9(8.9) 	 & $-$37.8(2.4) \\
IR Industries BB & 2016 Apr 21 	 & Clear	& 230.7(5.9)   	 & $-$122.0(4.4) 	 & $-$71.8(4.8) 	 & 260.9(5.6) 	 & $-$13.93(52) \\
Mikron BB  & 2016 Apr 22 	 & $B$	& $-$16(15)    	 & 117(14)       	 & $-$73(29)    	 & 118(14)    	 & 48.9(3.5)    \\
Mikron BB  & 2016 Apr 22 	 & $V$	& $-$50.4(4.3) 	 & 84.4(4.6)     	 & $-$70(11)    	 & 98.2(4.5)  	 & 60.4(1.3)    \\
\hline
8" IS\tablenotemark{b} + IB\tablenotemark{c}      & 2016 Apr 19	& $B$ 	 & 406.9(2.5) 	 & $-$181.5(2.0) 	 & 9.9(4.0)     	 & 445.5(2.4) 	 & $-$12.02(14) \\
12" IS + IB      & 2016 Apr 19	& $B$ 	 & 137.0(4.2) 	 & $-$93.9(4.2)  	 & $-$84.7(8.9) 	 & 166.0(4.2) 	 & $-$17.22(73) \\
12" IS + IB & 2016 Apr 21	& $U$ 	 & 247(20)    	 & 22(20)        	 & $-$306(50)   	 & 247(20)    	 & 2.6(2.3)     \\
12" IS + IB & 2016 Apr 21	& $B$ 	 & 214.9(4.4) 	 & 12.0(4.4)     	 & $-$74.0(9.2) 	 & 215.2(4.4) 	 & 1.60(59)     \\
12" IS + IB & 2016 Apr 21	& $V$ 	 & 178.3(2.8) 	 & 15.5(3.1)     	 & $-$78.5(7.7) 	 & 178.9(2.8) 	 & 2.49(50) \\
12" IS + IB & 2016 Apr 21	& Clear 	 & 179.5(2.6) 	 & 18.2(1.9)     	 & $-$41.1(2.1) 	 & 180.4(2.6) 	 & 2.89(31) \\
\hline
FS\tablenotemark{d} + lens  & 2018 Jul 16	& $B$ 	 & 63(30) 	 & $-$260(44) 	 & 85(96)     	 & 266(43) 	 & 141.8(3.4) \\
FS + lens  & 2018 Jul 16 	& Clear	 & 86(20) 	 & $-$228(20) 	 & $-$143(22) 	 & 243(20) 	 & 145.4(2.3) \\
FS + lens & 2018 Jul 16 	& Mean	 & 82(23) 	 & $-$233(18) 	 & $-$132(22) 	 & 246(19) 	 & 144.7(2.6) \\
\hline
\hline
Source 	& UT Date		 & Band	 & $q$ (\%)		& $u$ (\%)		& $v$ (\%)		 & $p$ (\%)		 & $\Theta$ $(^\circ)$ \\
\hline
IB on-axis & 2016 Apr 15	& $B$ 	 & 0.5310(10)    	 & 0.26177(94) 	 & 0.01913(49)    	 & 0.5920(10) 	 & 13.121(46) \\
IB off-axis & 2016 Apr 15	& $B$ 	 & 2.2411(53)    	 & 1.3606(48)  	 & 0.0240(18)     	 & 2.6218(52) 	 & 15.632(54) \\
Leica & 2016 Apr 15	& $B$ 	 & $-$3.7021(62) 	 	& 2.946(11)   	 & $-$0.13465(82) 	& 4.7313(83) 	 & 70.744(56) 
\label{tab:labsrc}
\enddata
\tablenotetext{a}{BB = blackbody}
\tablenotetext{b}{IS = integrating sphere}
\tablenotetext{c}{IB = incandescent bulb}
\tablenotetext{d}{FS = fiber source}
\end{deluxetable*}

We inject a variety of light sources into POLISH2 to determine the most unpolarized, man-made source to calibrate the POLISH2 polarization zero point in the lab (Table \ref{tab:labsrc}). In the 2016 measurements, all sources flood-illuminated the POLISH2 aperture. Here, two cavity blackbodies (Mikron Infrared M360 and one from IR Industries) were set to 1263 and 1173 K, respectively, and injected into POLISH2. Additionally, two sizes of Labsphere integrating spheres were used (8" and 12" diameter) with a desk lamp and incandescent bulb as the light source. Finally, both the incandescent desk lamp and a Leica lamp with a focusing lens were injected directly into POLISH2. Data were obtained with the long axis of the incandescent bulb pointed first at the POLISH2 aperture and then $90^\circ$ away from the POLISH2 aperture. In the 2018 measurements, a Cole-Parmer fiber light source illuminated a 1 mm diameter pinhole, which was reimaged by an $f/16$ lens to focus on the POLISH2 field stops and field lenses. This $f/$ratio mimics that experienced at Gemini North, the Lick 3-m, and the Lick 1-m.

As perhaps expected, both cavity blackbodies deliver the most unpolarized beam to POLISH2. However, the linear (detected with $3.5 < $ SNR $< 47$) and circular polarization (detected with $2.5 < $ SNR $< 15$) of both blackbodies is measured to lie at the 100 ppm = 0.01\% level, which is one to two orders of magnitude larger than unpolarized stars. This non-zero linear polarization must be intrinsic to each blackbody, and not simply due to some instrumental offset in POLISH2, because polarization orientation $\Theta$ is similar from band to band for each blackbody but varies significantly between blackbodies. That is, two truly unpolarized blackbodies should appear identical to POLISH2, and any instrumental offset should be the same from blackbody to blackbody.   

The significantly larger Mikron blackbody delivers about half the circular polarization of the smaller IR Industries blackbody, which suggests that non-zero circular polarization is also intrinsic to the blackbodies. However, both the sign of circular polarization and the degree of linear polarization are similar between blackbodies. No significant variation in blackbody polarization (linear or circular) is measured between $B$ and $V$ bands for each blackbody, though the beam from the IR Industries blackbody has enhanced linear polarization and reduced circular polarization through the clear bandpass.

The next more polarized lab source studied is an incandescent bulb illuminating two integrating spheres. Linear polarization is detected with $12 < $ SNR $< 190$, and circular polarization is detected with $2.4 < $ SNR $< 19$. While linear polarization of the 8" diameter sphere is larger than that of the 12" diameter sphere, their polarization orientations are similar. Polarization orientation changes by $\sim 20^\circ$ in the 2.8 days between runs, which may be caused by an unidentified but slight change to the lab setup. Interestingly, circular polarization of the 8" sphere is consistent with zero, even though its linear polarization is nearly three times that of the 12" sphere. This reduction of circular polarization given an enhancement in linear polarization is similar to that observed in the IR Industries blackbody between the $B/V$ and clear bandpasses. Circular polarization of the 12" sphere is consistent with that of the Mikron blackbody but half that of the IR Industries blackbody. Circular polarization from the 12" sphere is negative like that from both blackbodies, which may suggest a common physical origin to circular polarization from cavity blackbodies and integrating spheres. Unlike the blackbodies, the 12" sphere's degree of linear and circular polarization both decrease with increasing wavelength. The absolute value of circular polarization decreases with increasing wavelength more quickly than does the degree of linear polarization.

Next, the Cole-Parmer fiber source generally delivers slightly larger degree of linear (detected with $6.2 < $ SNR $< 13$) and circular polarization (detected with $0.9 < $ SNR $< 6.5$) than the integrating spheres. Like the integrating spheres, the fiber source delivers weaker circular polarization than the IR Industries blackbody, though the difference is slight for the fiber source. While $B$ and clear bandpass Stokes $v$ measurements differ by only $2.3 \sigma$, it is instructive to note that their signs are different. Recall that Stokes parameter value is given by the magnitude of the $(X,Y)$ phasor from the lock-in amplifier reduction code (Equation \ref{eq:stokesval}), while sign is given by its phase $\Phi$ (Equation \ref{eq:stokessign}). POLISH2's normally distributed, observable quantities are $X$ and $Y$, not Stokes parameters. Thus, Stokes parameter measurements with zero mean will actually follow a bimodal distribution, because normally distributed $X$ and $Y$ will generate both positive and negative signs to individual Stokes parameter measurements. That is, the Stokes parameter value is the positive definite, quadrature sum of $X$ and $Y$ that may later have a negative sign applied to it. However, the bias inherent in a quadrature sum means that no Stokes parameter measurement will ever truly be zero. In the case of the $B$ band Stokes $v$ measurement of the fiber source in Table \ref{tab:labsrc}, its true sign may be negative like that of the clear bandpass measurement, even though data reduction suggests its sign to be positive.

The last set of lab tests to be discussed here involve a bare, incandescent bulb and a Leica source with a focusing lens injected directly into POLISH2. These sources are strongly linearly polarized (detected with $510 < $ SNR $< 590$). When the long axis of the incandescent bulb is pointed into the POLISH2 aperture (``on-axis" in Table \ref{tab:labsrc}), linear polarization of $p \sim 0.6\%$ and circular polarization of $v \sim 0.02\%$ (detected with SNR = 39) are measured. When the long axis of the bulb is rotated $90^\circ$ away from the POLISH2 line of sight (``off-axis" in Table \ref{tab:labsrc}), a significant amount of light from the filament is scattered into the POLISH2 aperture by the section of the bulb along the its long axis. This scattering geometry increases linear polarization by a factor of 4.4 (to $p \sim 2.6\%$) but has very little effect on the polarization orientation. It also only increases circular polarization by a factor of 1.3 (still at $v \sim 0.02\%$ and detected with SNR = 14). Interestingly, while both cavity blackbodies, integrating spheres, and fiber source measurements emphatically have zero or negative circular polarization, both incandescent bulb geometries generate highly significant, positive circular polarization. This suggests that the incandescent bulb coating may convert linear polarization to circular polarization with a different handedness. Finally, the Leica focusing source is dramatically polarized, with $p \sim 4.7\%$ and $v \sim -0.1\%$ (detected with SNR = 160). The polarization orientation is significantly different from that of the incandescent bulb ($\Delta\theta \sim 56^\circ$). It seems reasonable that the strong linear and circular polarization of this source is due to stress birefringence in the highly curved, short focal length focusing lens. Its negative circular polarization is consistent with all but the incandescent bulb, which may provide clues as to how the polarization from these sources is generated.

\subsection{On-Sky Measurements}
\label{sec:skycal}
\subsubsection{Gemini North 8-m Telescope Polarization}
\label{sec:gem}

As mentioned in section \ref{sec:altaz}, the alt-az nature of Gemini North enables calibration of Stokes $q$ and $u$ modulation efficiencies as telescope polarization rotates with parallactic angle. For example, were telescope polarization to be aligned with instrumental Stokes $+q$ at a parallactic angle of $0^\circ$, it must then be aligned with instrumental Stokes $+u$ at a parallactic angle of $45^\circ$. Thus, to first order, the amplitude of the sinusoidal variations of Stokes $q$ and $u$ with parallactic angle must be identical, since the degree of telescope polarization is invariant under rotation.

During the UT 11 to 14 November 2016 Gemini North commissioning of POLISH2, we observed the hot Jupiter host star WASP-12 \citep[orbital period $\sim 1.09$ d,][]{Hebb2009}, the nearly unpolarized calibrators HR 1791 and Neptune, the strongly linearly polarized calibrators 53 Per and 55 Cyg, the long-period eclipsing binary EE Cep, the pulsating white dwarf G29-38, and the Crab Pulsar in $B$ band with Hamamatsu H10721-110 SEL PMTs. During the Gemini North POLISH2 run on UT 2 to 7 August 2018, we observed the hot Jupiter host star HD 189733 \citep[orbital period $\sim 2.22$ d,][]{Bouchy2005}, the strongly linearly polarized HD 204827, the strongly circularly polarized white dwarf Grw $+70^\circ8247$ \citep{Kemp1970}, and the nearly unpolarized calibrator star HR 8585 in $B$ band with H10721-210 SEL PMTs. HR 8585 was also observed in a clear, unfiltered bandpass to calibrate for telescope polarization in support of observations of solar system asteroids. The PMTs used on the 2018 POLISH2 run have a slightly improved QE at all wavelengths with respect to those used during the 2016 POLISH2 commissioning run, but they are otherwise nearly identical.

We limit the discussion of exoplanet host polarization to time-averaged measurements of intrinsic stellar polarization, and we leave discussion of any exoplanet modulation or stellar variability to a future publication. WASP-12 was observed over five pointings of roughly 45 min duration on each of the three nights of the 2016 run, and its unpolarized calibrator HR 1791 was observed over five pointings of roughly 5 min duration before each WASP-12 pointing. This enabled a large range in parallactic angle to be obtained for the short-duration HR 1791 observations to maximize the accuracy of the telescope and stellar polarization fit. Overhead due to slewing and acquisition took 3 min. A similar cadence of observations was obtained on HD 189733, HD 204827, and HR 8585 in 2018.

\begin{figure*}
\centering
\includegraphics[width=0.7\textwidth]{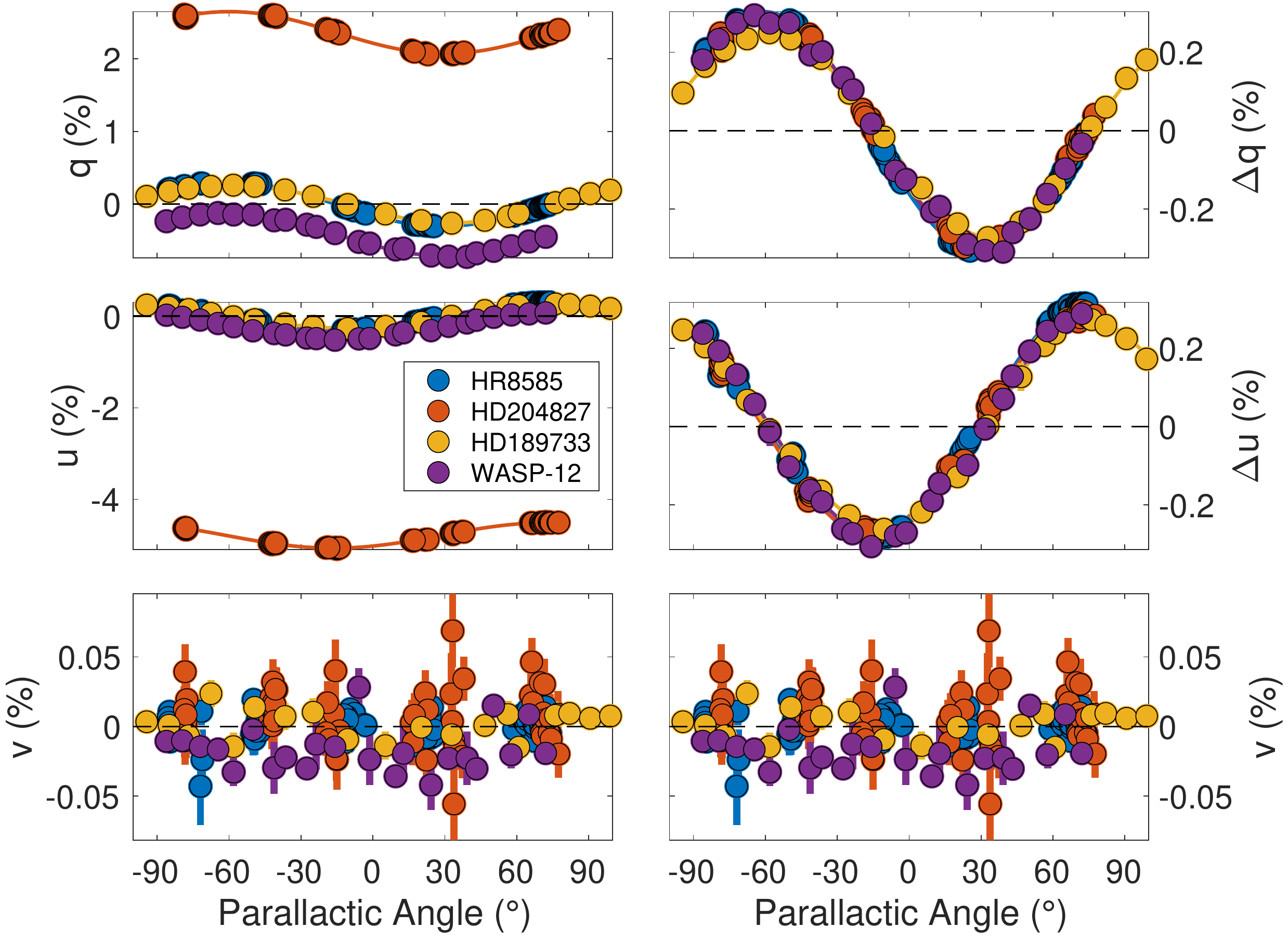}
\caption{Gemini North POLISH2 $B$ band measurements of the exoplanet hosts HD 189733 and WASP-12, the nearly unpolarized HR 8585, and the strongly polarized HD 204827. WASP-12 was observed on 12 to 14 November 2016 UT, while the others were observed on 2 to 7 August 2018 UT. Data are aggregated over all nights of each run and fit by Equations \ref{eq:gemq} and \ref{eq:gemu}. \textit{Left:} Intrinsic stellar polarization $q_*$ (top) and $u_*$ (middle) are included, which manifest as constant offsets with respect to parallactic angle. \textit{Right:} Stellar polarization is removed to isolate telescope polarization. Intrinsic and telescope-induced circular polarization are conflated (bottom). Note the change in scale in the bottom panel.}
\label{fig:gemini}
\end{figure*}

All stars save HR 1791 were observed with the Gemini de-rotator in Follow mode, where field de-rotation is enabled and instrumental Stokes $+Q$ is aligned with Celestial Stokes $+Q$ at all times. In this mode, telescope polarization rotates with parallactic angle $\phi$. Figure \ref{fig:gemini} shows $B$ band measurements of the above stars as a function of parallactic angle, which are fit by second order Fourier series:
\begin{subequations}
\begin{align}
\label{eq:gemq}
q(\phi) &= q_* + a_\text{TP} \cos \omega \phi + b_\text{TP} \sin \omega \phi \\
\nonumber & + a_2 \cos 2\omega \phi + b_2 sin 2\omega \phi \\
\label{eq:gemu}
u(\phi) &= u_* + c_\text{TP} \cos \omega \phi + d_\text{TP} \sin \omega \phi \\
\nonumber & + c_2 \cos 2\omega \phi + d_2 sin 2\omega \phi \\
\label{eq:gemv}
v(\phi) & = v_* + v_\text{TP} \\
\label{eq:tpq}
\text{TP}_q & = \sqrt{a_\text{TP}^2 + b_\text{TP}^2} \\
\label{eq:tpu}
& = \sqrt{c_\text{TP}^2 + d_\text{TP}^2} = \text{TP}_u.
\end{align}
\end{subequations}

Here, stellar polarization $q_*$ and $u_*$ are constant with respect to parallactic angle. Figure \ref{fig:gemini} illustrates the complementary effects caused by stellar and telescope polarization. Telescope polarization is given by the quadrature sum of the first order Fourier components (Equations \ref{eq:tpq} and \ref{eq:tpu}). Note that the telescope polarization measured by Stokes $q$ should be identical to that measured by Stokes $u$, because Stokes $q$ and $u$ swap after rotation of parallactic angle by $45^\circ$. For all but HD 189733 observations, the second order Fourier components are consistent with zero. Tables \ref{tab:gemtpdat} and \ref{tab:gemdat} list telescope and stellar polarization fits, respectively.

Since circular polarization is invariant under rotation, stellar and telescope-induced circular polarization are conflated (Equation \ref{eq:gemv} and Table \ref{tab:gemtpdat}). Thus, circular polarimetry at both alt-az and equatorial telescopes requires observations of nearly unpolarized stars. Telescope and stellar circular polarization measured on HR 8585 is consistent between $B$ and clear bandpasses, so we assume telescope circular polarization to be the weighted mean of measurements in these two bandpasses. Circular polarization measured on other stars is subtracted by this weighted mean, and intrinsic stellar circular polarization is listed in Table \ref{tab:gemdat}.

\begin{deluxetable*}{ccccccccc}
\tabletypesize{\normalsize}
\tablecaption{Gemini North POLISH2 Telescope Polarization Fits}
\tablewidth{0pt}
\tablehead{
\colhead{Star} & \colhead{UT Dates}& \colhead{Band} & \colhead{$\text{TP}_q (\%)$} & \colhead{$\text{TP}_u (\%)$} & \colhead{$\overline{\text{TP}}_{qu} (\%)$} & \colhead{$\text{TP}_v + v_*$ (ppm)}}
\startdata
HR8585   & 2018 Aug 2$-$7  	& $B$ 	 & 0.30515(58) 	 & 0.29807(46) 	 & 0.3008(34)	& 21.4(5.4) \\
HR8585   & 2018 Aug 2$-$6  	& Clear 	 & 0.27045(63) 	 & 0.23715(52) 	 & 0.251(16)	& 24.1(3.2)  \\
HR8585   & $-$  			& Mean 	 & $-$	 	 & $-$	  	 & $-$	 	& 23.4(1.2)  \\
\hline
HD189733 & 2018 Aug 2$-$7  	& $B$ 	 & 0.2572(19)  	 & 0.2644(14)  	 & 0.2620(34) 	 & 37.6(9.0) \\
\hline
WASP-12  & 2016 Nov 12  	& $B$                      	 & 0.2945(50) 	 & 0.2962(40) 	 & 0.29554(87) & $-$111(37)  \\
WASP-12  & 2016 Nov 13 	& $B$                       	 & 0.3006(41) 	 & 0.2939(79) 	 & 0.2992(27) 	 & $-$148(38)  \\
WASP-12  & 2016 Nov 14 	& $B$                       	 & 0.2861(64) 	 & 0.2851(47) 	 & 0.28544(48) 	 & $-$79(41) \\
WASP-12\tablenotemark{a}  & 2016 Nov 12$-$14 	& $B$ 	 & 0.2992(33) 	 & 0.2852(27) 	 & 0.2908(69) 	 & $-$115(22)  \\
WASP-12\tablenotemark{b}  & 2016 Nov 12$-$14 	& $B$ 	 & 0.2973(21) 	 & 0.2883(20) 	 & 0.2881(46) 	 & $-$115(16)    \\
\hline
HD204827 & 2018 Aug 2$-$7  	& $B$ 	 & 0.2855(18)  	 & 0.2763(20)  	 & 0.2813(46)	& 50(87)
\label{tab:gemtpdat}
\enddata
\tablenotetext{a}{Fit to aggregated data from all nights}
\tablenotetext{b}{Weighted mean of fits from individual nights}
\end{deluxetable*}

\begin{deluxetable*}{ccccccccc}
\tabletypesize{\normalsize}
\tablecaption{Gemini North POLISH2 Stellar Polarization Fits}
\tablewidth{0pt}
\tablehead{
\colhead{Star} & \colhead{$d$ (pc)} & \colhead{Band} & \colhead{$q_*$ (ppm)} & \colhead{$u_*$ (ppm)} & \colhead{$v_*$ (ppm)} & \colhead{$p_*$ (ppm)} & \colhead{$\Theta_* (^\circ)$}}
\startdata
HR8585   & 31.46(12)	& $B$ 	 & $-$2.3(3.7)    & $-$12.1(3.6)   & $-$2.0(5.4)    	 & 11.7(3.6)	 & 129.5(9.1) \\
HR8585   & $\cdots$		& Clear 	 & $-$2.6(4.4)	 & 8.1(4.0) 	 & 0.7(3.2)    	 & 7.4(4.0) 	 & 54(16)   \\
HD189733 & 19.7758(63)	& $B$ 	 & $-$89(13)     	 & 32.7(9.6) 	 & 14.3(9.0) 	 & 94(13)  	 	& 79.9(3.1)    \\
\hline
\hline
Star		& $d$ (pc)		& Band	& $q_* (\%)$		& $u_* (\%)$ 		& $v_* (\%)$ 		& $p_* (\%)$ 		& $\Theta_* (^\circ)$ \\
\hline
WASP-12\tablenotemark{a}  & 413.0(2.8)	& $B$ 	 & 0.4133(34) 	 & 0.2081(30) 	 & $-$0.0135(37) 	 & 0.4628(33) 	 & 13.36(19)  \\
WASP-12\tablenotemark{b}  & $\cdots$	& $B$   	 & 0.4163(29) 	 & 0.2189(49) 	 & $-$0.0172(38) 	 & 0.4703(34) 	 & 13.87(28)  \\
WASP-12\tablenotemark{c}  & $\cdots$	& $B$   	 & 0.4059(52) 	 & 0.2006(31) 	 & $-$0.0102(41) 	 & 0.4528(48) 	 & 13.15(23)  \\
WASP-12\tablenotemark{d}  & $\cdots$	& $B$   	 & 0.4151(15) 	 & 0.2042(14) 	 & $-$0.0138(16) 	 & 0.4626(15) 	 & 13.100(87)   \\
WASP-12\tablenotemark{e}  & $\cdots$	& $B$   	 & 0.4169(23) 	 & 0.2017(20) 	 & $-$0.0138(22) 	 & 0.4631(22) 	 & 12.91(12)  \\
HD204827 & 929(85)	& $B$ 	 & $-$2.3574(13)     	 & 4.7856(14)  	 & 0.0026(87)      	 & 5.3347(14)  	 & 58.1123(69)    
\label{tab:gemdat}
\enddata
\tablenotetext{a}{UT 12 Nov 2016}
\tablenotetext{b}{UT 13 Nov 2016}
\tablenotetext{c}{UT 14 Nov 2016}
\tablenotetext{d}{Fit to aggregated data from all nights}
\tablenotetext{e}{Weighted mean of fits from individual nights}
\end{deluxetable*}

Figure \ref{fig:gemini} and Table \ref{tab:gemtpdat} show that telescope linear polarization at Gemini North is $\sim 0.3\%$, which is an order of magnitude larger than that measured at 2.5 to 5-m telescopes (section \ref{sec:altaz}). Private communication with Gemini Observatory staff (Andrew Adamson, Thomas Hayward, and Thomas Schneider) reveal that the primary mirror M1 is coated by a magnetron as the mirror rotates about its optical axis, while the secondary mirror M2 is coated during a single linear pass. Due to a misalignment in coating prior to POLISH2 observations, one side of M2 received a slightly thicker coating of NiCr than the other. This misalignment was corrected in 2021, subsequent to POLISH2 observations. Given that each Angstrom of NiCr coating causes $\sim 1\%$ loss in reflectivity in the visible, it seems reasonable that an asymmetric coating would also generate enhanced telescope polarization.

Regardless, telescope polarization of $\sim 0.3\%$ is broadly consistent between our November 2016 and August 2018 runs, and it is also consistent with HIPPI-2 measurements obtained in July 2018 \citep{Bailey2020, Cotton2020G29}, which suggests that it is fairly constant over time. We measure a statistically significant difference in telescope polarization from star to star at the $0.01\%$ level, which may be due to slight changes in effective bandpass from star to star coupled with the known, extreme variation of telescope polarization with wavelength. This variation with wavelength leads the HIPPI-2 team to conclude that polarimetric accuracy at Gemini North is limited to $\sim 25$ ppm \citep{Bailey2020}. It seems reasonable that this limitation may have resolved with the realignment of the mirror coating magnetron.

\begin{figure}
\centering
\includegraphics[width=0.47\textwidth]{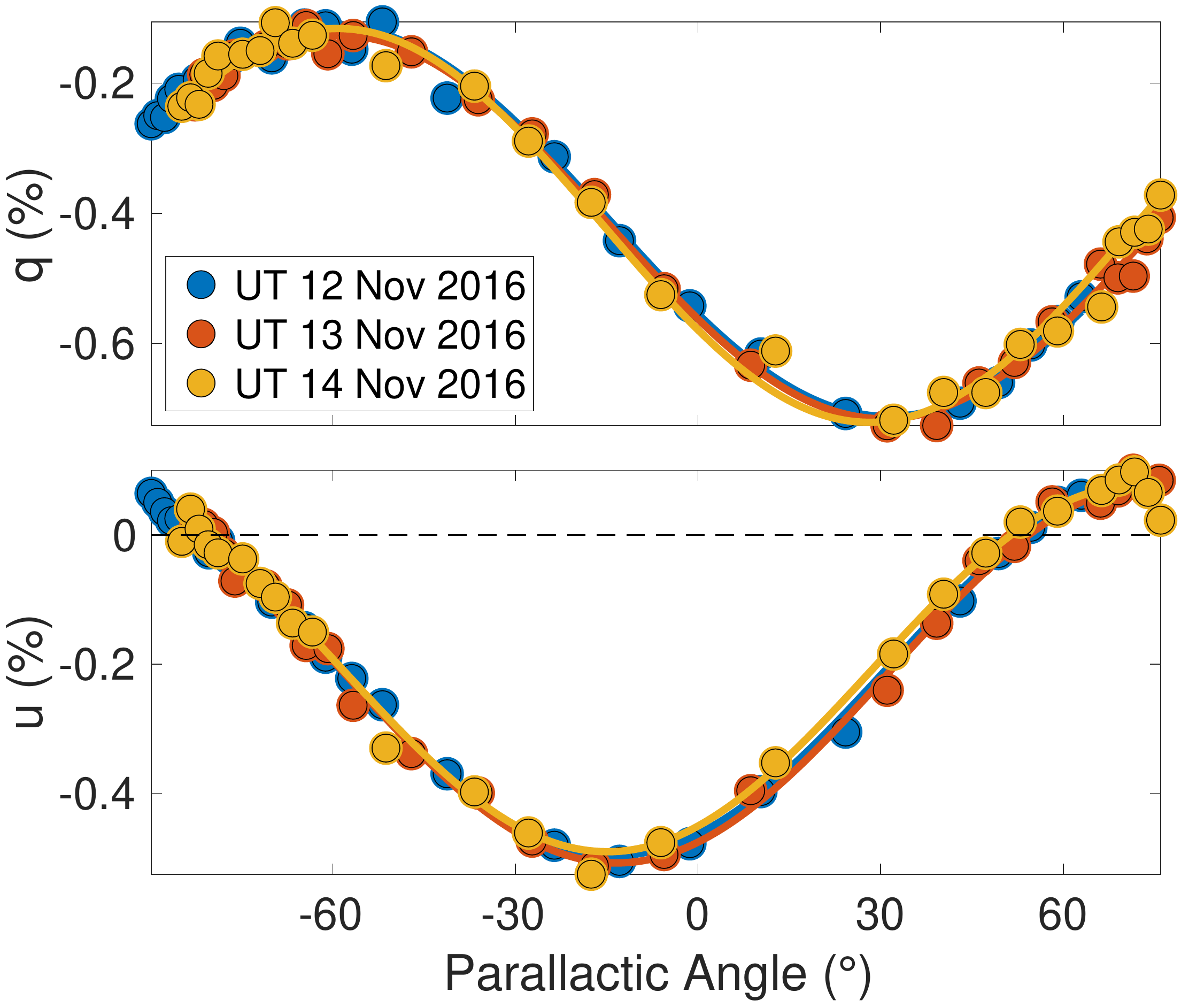}
\caption{Gemini North POLISH2 $B$ band measurements of the exoplanet host WASP-12 on three successive nights, which highlights the repeatability of self-calibration of telescope polarization.}
\label{fig:wasp12}
\end{figure}

However, Figures \ref{fig:gemini} and \ref{fig:wasp12} suggest that any star, if well sampled in parallactic angle, may enable accurate self-calibration of telescope polarization. Indeed, while the standard equatorial telescope procedure of observing nearly unpolarized stars to calibrate for science target polarization is clearly inadequate at Gemini North (Table \ref{tab:gemtpdat}), self-calibration of the science target itself at Gemini North is viable. This requires that any intrinsic temporal variability in the science target be averaged out with observations obtained at other timescales. For short-period exoplanet hosts such as HD 189733 and WASP-12, this requires observations on successive nights. Figure \ref{fig:wasp12} illustrates how such observations of WASP-12 enable consistent measurement of both stellar and telescope polarization. Indeed, Tables \ref{tab:gemtpdat} and \ref{tab:gemdat} demonstrate that nightly values of stellar and telescope polarization are consistent with each other, and the weighted mean of nightly fits to stellar and telescope polarization are consistent with the fits obtained when data from all nights are included in the fit.

We therefore conclude that the uncertainties in Table \ref{tab:gemdat}, many lying well below 25 ppm, demonstrate measurement accuracy at this level. We define accuracy as the ability to measure intrinsic variation in each star in spite of calibration artifacts. Indeed, observations such as those of HR 8585 in $B$ band, where $p = 11.7 \pm 3.6$ ppm, achieve such high accuracy because the alt-az Gemini North enables separation of intrinsic stellar polarization from that of the telescope. This powerful ability to separate stellar from telescope polarization is independent of stellar polarization itself, and it only scales with the number of photons detected. For example, observations of both the weakly polarized HD 189733 ($p \sim 0.0094\%$) and the strongly polarized HD 204827 ($p \sim 5.3\%$) achieve similar accuracy of telescope-subtracted linear polarization (13 and 14 ppm, respectively) because their brightnesses ($B \sim 8.6$ and 8.7, respectively) and observed range of parallactic angles are similar. As can be seen in Table \ref{tab:gemdat}, HR 8585 appears to be nearly unpolarized in both $B$ band and the clear bandpass. Given HR 8585's distance, this is consistent with results from the interstellar polarization study of \cite{Cotton2017}.

\subsubsection{Lick Observatory Telescope Polarization}
\label{sec:lick}

Previous POLISH/POLISH2 publications \citep{Wiktorowicz2008, Wiktorowicz2009, WiktorowiczNofi2015, Wiktorowicz2015_189} assumed that telescope polarization varied on a nightly basis. Observations of the brightest, most weakly polarized stars were used to estimate this nightly variability in telescope polarization. With the more rigorous calibration methodology presented in this paper, we revisit the assumption of nightly telescope polarization variability with POLISH2 at the Lick 3-m from 2011 to 2021 and at the Lick 1-m from 2011 to 2014.

Truly unpolarized stars will manifest with polarization of order 1 ppm ($p \sim 0.0001\%$) or larger due to photon noise, but even weakly polarized stars will be 100 times as polarized ($p \sim 100$ ppm, or 0.01\%). Rather than removing the latter stars from the calculation of telescope polarization, we collate observations of all stars with polarization at this level or below in the entire POLISH2 database. We then assume that telescope polarization is constant within each run, which is typically seven nights long. Weakly polarized stars with observations on at least two runs from 2011 to 2021 are selected, and each star's weighted mean, historical polarization observables $X_{QUV}/DC$ and $Y_{QUV}/DC$ in each of Stokes $q$, $u$, and $v$ (Equations \ref{eq:xydc} and \ref{eq:quv}) are calculated and subtracted. This removes intrinsic stellar polarization so run-to-run variations in telescope polarization may be uncovered.

Next, telescope polarization variations from run to run are calculated from the weighted mean $X_{QUV}/DC$ and $Y_{QUV}/DC$ of all mean-subtracted stellar observations in each run. Since input stellar data to this step are mean-subtracted, this step only provides relative telescope polarization variability and not absolute, time-averaged telescope polarization values. To determine time-averaged telescope polarization, historical stellar data are subtracted by the estimate of telescope polarization variability during each run, which leaves stars to cluster at non-zero polarization values that no longer vary from run to run. Thus, time-averaged telescope polarization is given by the weighted mean polarization of these clusters in $X_{QUV}/DC$ and $Y_{QUV}/DC$. Here, final, absolute telescope polarization from run to run is given by the sum of relative, run-to-run telescope polarization variability and absolute, time-averaged polarization values. Science data are subtracted by this look-up table to correct for telescope polarization. For runs without observations obtained in a certain band, piecewise cubic Hermite interpolating polynomials are used to interpolate $X_{QUV}/DC$ and $Y_{QUV}/DC$ at the mean date of the runs.

\begin{figure}
\centering
\includegraphics[width=0.47\textwidth]{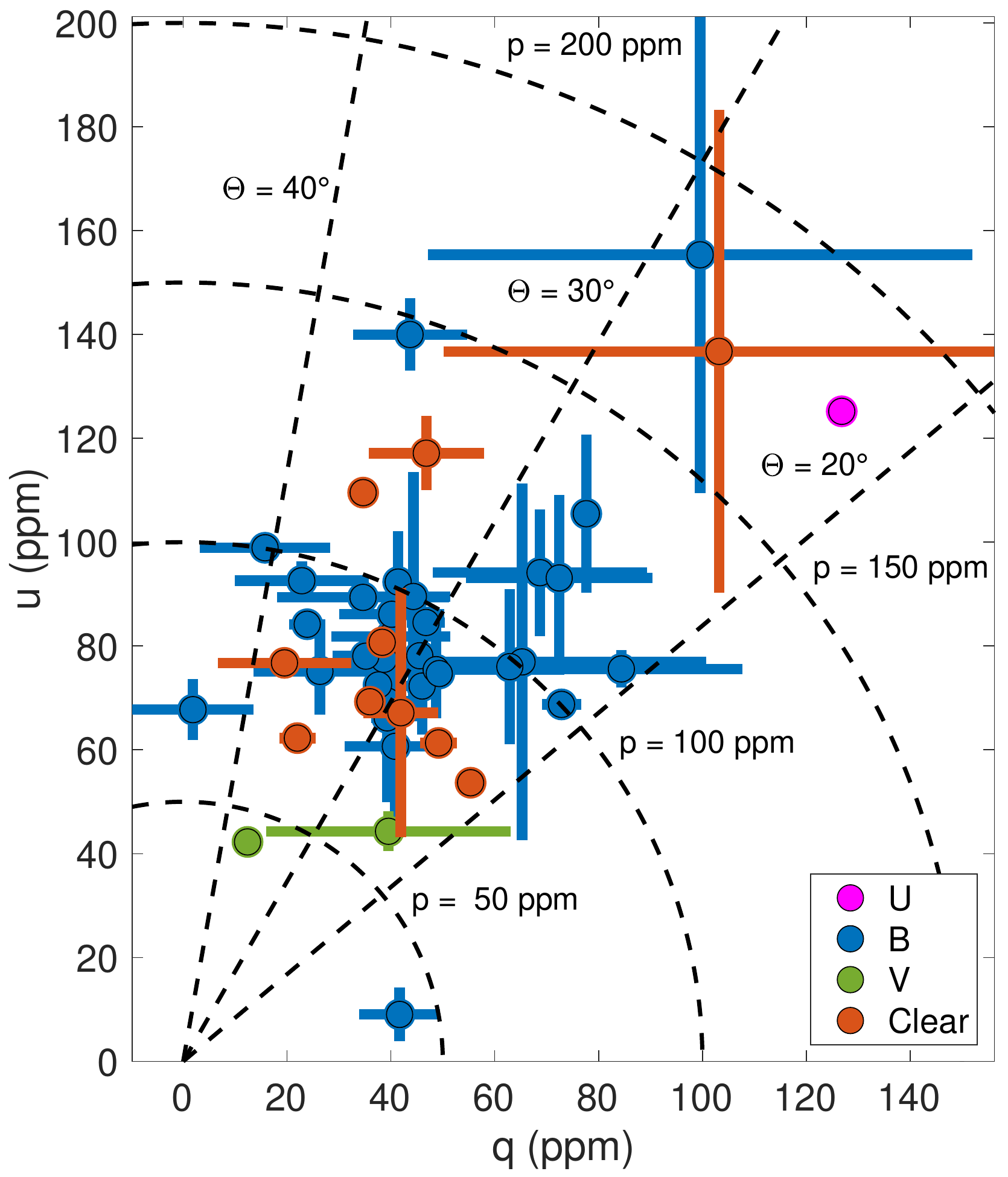}
\caption{Lick 3-m POLISH2 $UBV$ and clear bandpass measurements of telescope polarization from 2011 to 2021. Telescope polarization is typically $p \sim 100$ ppm (0.01\%) with an orientation of $\Theta \sim 30^\circ$.}
\label{fig:licktp}
\end{figure}

\begin{figure}
\centering
\includegraphics[width=0.47\textwidth]{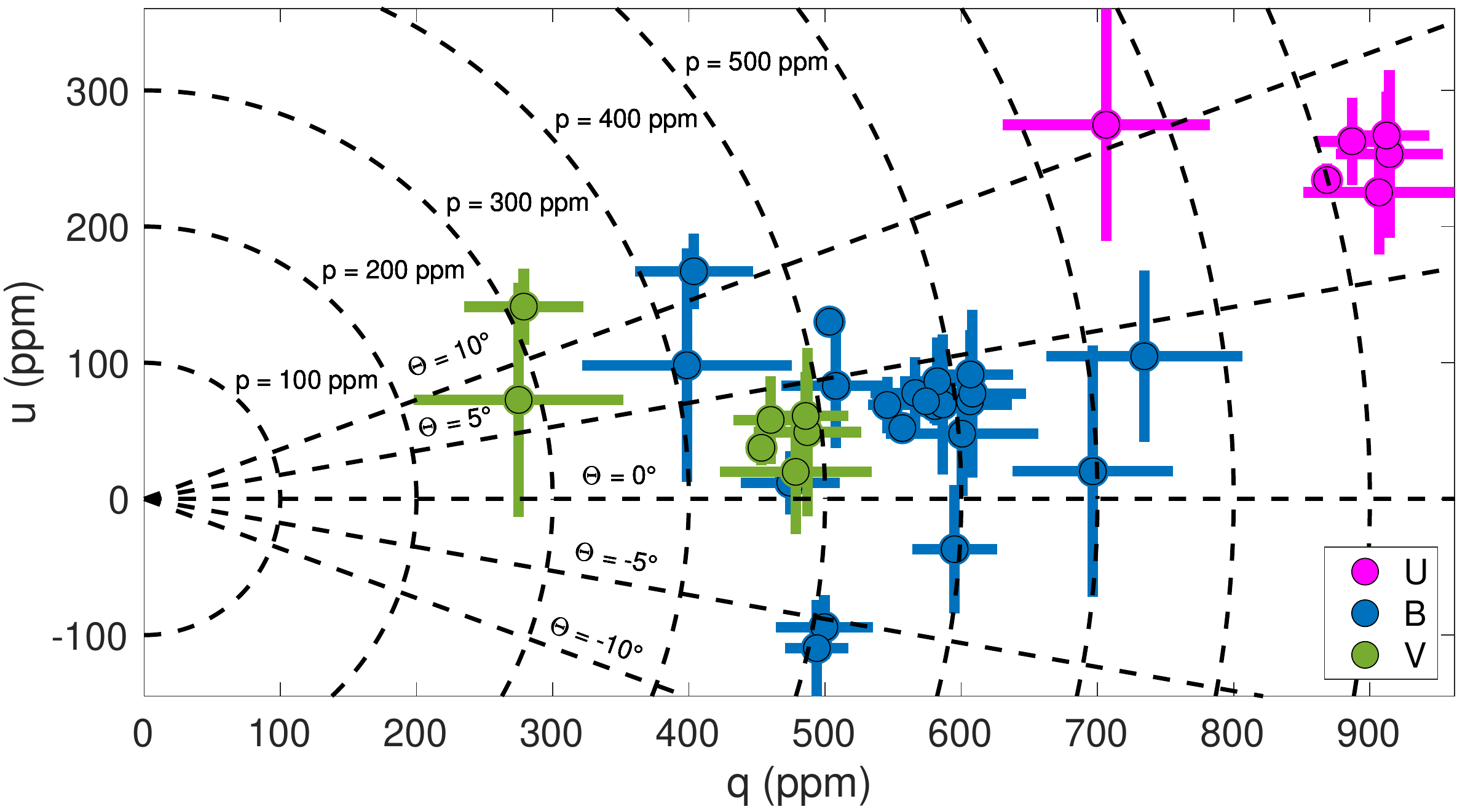}
\caption{Lick 1-m POLISH2 $UBV$ measurements of telescope polarization from 2011 to 2014. Telescope polarization is typically $p \sim 500$ ppm (0.05\%) with an orientation of $\Theta \sim 0^\circ$ with respect to Celestial North. Polarization data are rotated by the $\Theta^\prime \sim 151^\circ$ rotational offset of the Lick 1-m Cassegrain ring (section \ref{sec:step2}).}
\label{fig:licktp1}
\end{figure}

Figure \ref{fig:licktp} and Table \ref{tab:licktp} illustrate telescope polarization converted to Stokes $q$, $u$, and $v$ for the Lick 3-m, while Lick 1-m data are shown in Figure \ref{fig:licktp1} and Table \ref{tab:licktp1}. Linear and circular polarization ratios across telescope and/or band are listed in Tables \ref{tab:licktplinratios} and \ref{tab:licktpcircratios}, respectively. It is clear that telescope polarization is strongly wavelength dependent at the Lick 3-m, as degree of polarization decreases by a factor of $3.987 \pm 0.096$ from $U$ to $V$ bands (Table \ref{tab:licktplinratios}). Given the breadth of the clear bandpass (Table \ref{tab:qe}), coupled with the use of blue-sensitive Hamamatsu H10721-210 SEL and red-sensitive H7422P-40 SEL PMTs, it is difficult to conclude anything definitive about telescope polarization in the clear bandpass. Chromaticity of Lick 1-m telescope polarization is half as severe as that of the Lick 3-m, as polarization decreases by a factor of $2.16 \pm 0.17$ from $U$ to $V$ bands. However, the Lick 1-m tends to have 5 to 10 times stronger linear polarization than the Lick 3-m in all bands.

Mean values in Tables \ref{tab:licktp} and \ref{tab:licktp1} indicate weighted mean Stokes $q$, $u$, and $v$ for observations in a given bandpass obtained across all observing runs at a given telescope (Equation \ref{eq:meanwt}). A $\chi^2$ test is performed to search for variability in each Stokes parameter. If the hypothesis of a constant Stokes parameter value can be rejected with at least $3 \sigma$ confidence, telescope polarization is assumed to be variable from run to run, and uncertainty in the weighted mean is given by the square root of the weighted variance $\sigma_\text{wv}$ (Equation \ref{eq:std}). This captures the amplitude of intrinsic variability. If not, the standard error of the weighted mean $\sigma_\text{se}$ is used and the data are assumed to be non-variable (Equation \ref{eq:stderr}):
\begin{subequations}
\begin{align}
 \label{eq:meanwt}
\bar{q} & = \frac{\smashoperator{\sum_{i=1}^{n}} \sigma_{q_i}^{-2} q_i}{\smashoperator{\sum_{i=1}^{n}} \sigma_{q_i}^{-2}} \\
 \label{eq:std}
\sigma_{\text{wv}_q} & = \sqrt{\frac{\smashoperator{\sum_{i=1}^{n}} \sigma_{q_i}^{-2}(q_i - \bar{q})^2}{\smashoperator{\sum_{i=1}^{n}} \sigma_{q_i}^{-2}}} \\
 \label{eq:stderr}
\sigma_{\text{se}_q} & = \sqrt{\frac{1}{\smashoperator{\sum_{i=1}^{n}} \sigma_{q_i}^{-2}}}.
\end{align}
\end{subequations}

\noindent From Table \ref{tab:licktp}, we conclude that the variability in Lick 3-m telescope polarization in the $B$, $V$, and clear bandpasses is 10 ppm or less. The Lick 3-m campaign consists of 31 runs across ten years in $B$ band, two runs separated by one year in $V$ band, and ten runs across six years in the clear bandpass. Thus, the calibration methodology discussed in this publication has halved the measured variability in telescope polarization with respect to \cite{Wiktorowicz2015_189}. Additionally, we have determined that telescope polarization may be assumed to be constant during each run. From Table \ref{tab:licktp1}, we conclude that the variability in Lick 1-m telescope polarization in $UBV$ bandpasses, which sample 22 runs across three years, is about 50 ppm or less.

Most telescope polarization observations at the Lick 3-m were obtained in $B$ band to calibrate the search for scattered light from hot Jupiters. The addition of telescope polarization observations obtained in a clear bandpass began in 2015 to calibrate clear observations of rotation phase-locked variations of polarization from Main Belt Asteroids as well as observations of the Crab Pulsar. We relegate these results to future publications. As with Lick 3-m observations, most Lick 1-m observations were also obtained in $B$ band to identify and verify the most suitable, weakly polarized calibrator stars to support the Lick 3-m exoplanet campaign.

\begin{deluxetable*}{ccccccc}
\tabletypesize{\small}
\tablecaption{Lick 3-m Telescope Polarization}
\tablewidth{0pt}
\tablehead{
\colhead{UT Date Range}	& \colhead{Band}	& \colhead{$q$}	& \colhead{$u$}	& \colhead{$v$}		& \colhead{$p$}	& \colhead{$\Theta$} \\
 					 & & (ppm) 			& (ppm)			& (ppm)			& (ppm)			& ($^\circ$)}
\startdata
2011 Jul 21$-$22    	 & $B$   	 & 41.41(95)  	 & 92.3(9.8)  	 & 4.6(7.5)               	 & 101.1(8.9) 	 & 32.9(1.2) \\
2011 Aug 13$-$17    	 & $B$   	 & 69(21)     	 & 94(12)     	 & 15(20)                 	 & 115(16)    	 & 26.9(4.5) \\
2011 Nov 5$-$9      	 & $B$   	 & 45.6(2.2)  	 & 78.1(2.6)  	 & 27(12)                 	 & 90.4(2.5)  	 & 29.87(73) \\
2011 Dec 3$-$7      	 & $B$   	 & 46.0(3.7)  	 & 72.3(9.4)  	 & 21.4(7.1)              	 & 85.5(8.1)  	 & 28.8(2.0) \\
2012 Feb 4$-$6      	 & $B$   	 & 72(18)     	 & 93(16)     	 & 23.1(4.9)              	 & 117(17)    	 & 26.1(4.2) \\
2012 Apr 3$-$10     	 & $B$   	 & 40(11)     	 & 81.9(5.7)  	 & 36(22)                 	 & 90.5(7.2)  	 & 32.0(3.3) \\
2012 May 5$-$9      	 & $B$   	 & 2(12)      	 & 67.7(5.9)  	 & 32(14)                 	 & 66.8(5.9)  	 & 44.2(5.0) \\
2012 Jun 7$-$12     	 & $B$   	 & 49(25)     	 & 75.4(9.4)  	 & 24.4(1.9)              	 & 87(16)     	 & 28.6(7.0) \\
2012 Dec 4$-$5      	 & $B$   	 & 65(35)     	 & 77(34)     	 & 4(19)                  	 & 95(35)     	 & 25(11)    \\
2013 Jan 29$-$Feb 1 	 & $B$   	 & 63(12)     	 & 76(15)     	 & 61(85)                 	 & 98(14)     	 & 25.2(3.9) \\
2013 Feb 2$-$4      	 & $B$   	 & 84(23)     	 & 75.6(3.6)  	 & 13(36)                 	 & 112(18)    	 & 20.9(4.1) \\
$\cdots$      	 	& $V$   	 & 40(24)     	 & 44.3(3.8)  	 & 94(36)                 	 & 57(16)     	 & 24.1(9.1) \\
2013 Apr 27$-$29    	 & $B$   	 & 40(10)     	 & 86.1(2.4)  	 & 41.9(8.7)              	 & 94.6(4.8)  	 & 32.5(2.8) \\
2013 May 24$-$31    	 & $B$   	 & 35(17)     	 & 89.4(1.6)  	 & 43.2(8.9)              	 & 94.6(6.2)  	 & 34.4(4.6) \\
2013 Aug 17$-$22    	 & $B$   	 & 77.6(2.3)  	 & 105(15)    	 & 67(32)                 	 & 131(12)    	 & 26.8(2.0) \\
2013 Sep 11$-$16    	 & $B$   	 & 40.9(9.8)  	 & 61(15)     	 & 83(22)                 	 & 72(14)     	 & 28.0(4.7) \\
2013 Oct 11$-$14    	 & $B$   	 & 39.3(3.1)  	 & 66(16)     	 & 10.4(5.6)              	 & 76(14)     	 & 29.6(3.3) \\
2014 Jan 4$-$7      	 & $B$   	 & 38.6(8.6)  	 & 77.44(30)  	 & 21(11)                 	 & 86.2(3.9)  	 & 31.7(2.6) \\
2014 Feb 14$-$20    	 & $B$   	 & 26(13)     	 & 75.1(8.4)  	 & 10(31)                 	 & 78.6(9.0)  	 & 35.3(4.5) \\
2014 Mar 8$-$19     	 & $B$   	 & 39.4(4.4)  	 & 69.4(4.5)  	 & 59.8(7.1)              	 & 79.7(4.4)  	 & 30.2(1.6) \\
2014 Apr 20$-$21    	 & $U$   	 & 126.8(1.0) 	 & 125.2(2.5) 	 & 71(13)                 	 & 178.2(1.9) 	 & 22.31(30) \\
$\cdots$		    	 & $B$   	 & 37.6(1.0)  	 & 72.5(2.6)  	 & 74.9(9.4)              	 & 81.7(2.4)  	 & 31.29(53) \\
$\cdots$		    	 & $V$   	 & 12.4(1.1)  	 & 42.3(2.6)  	 & 79(12)                 	 & 44.0(2.6)  	 & 36.83(83) \\
2014 Jun 7$-$13     	 & $B$   	 & 49.3(1.4)  	 & 74.70(36)  	 & 86(25)                 	 & 89.50(81)  	 & 28.28(37) \\
2014 Jul 10$-$13    	 & $B$   	 & 35.2(6.3)  	 & 78.05(72)  	 & 83.3(7.6)              	 & 85.4(2.7)  	 & 32.9(1.9) \\
2015 Feb 6$-$13     	 & Clear 	 & 55($-$)		 & 54($-$)		 & 40($-$)			 & 77($-$)		 & 22($-$)     \\
2015 Jul 25$-$31    	 & $B$   	 & 72.9(3.8)  	 & 68.75(39)  	 & 26(24)                 	 & 100.1(2.8) 	 & 21.67(74) \\
2015 Sep 22$-$29    	 & $B$   	 & 41.7(7.8)  	 & 9.0(5.1)   	 & 110(190) 	 & 42.3(7.7)  	 & 6.1(3.6)  \\
2017 Sep 7$-$13     	 & $B$   	 & 23.9(3.5)  	 & 84.19(18)  	 & 136.8(3.2)             	 & 87.45(97)  	 & 37.1(1.1) \\
$\cdots$		     	 & Clear 	 & 22.0(3.5)  	 & 62.26(21)  	 & 115.9(3.5)             	 & 65.9(1.2)  	 & 35.3(1.4) \\
2018 Jun 1$-$6      	 & Clear 	 & 35($-$)	 	& 110($-$)	 	& 51($-$)	 		& 115($-$)	 	& 36($-$) \\
2018 Sep 18$-$24    	 & $B$   	 & 44(11)     	 & 140.0(7.0) 	 & 32(24)                 	 & 146.3(7.4) 	 & 36.3(2.1) \\
$\cdots$		    	 & Clear 	 & 47(11)     	 & 117.1(7.1) 	 & 42(23)                 	 & 125.7(7.8) 	 & 34.1(2.4) \\
2019 May 18$-$23     & Clear 	 & 38($-$)	 	&  81($-$)	 	& 56($-$)	 		&  89($-$)	 	& 32($-$) \\
2019 Aug 10$-$16    	 & $B$   	 & 23(13)     	 & 92.6(3.8)  	 & 26.4(8.5)              	 & 94.5(4.8)  	 & 38.1(3.8) \\
2019 Oct 25$-$29    	 & $B$   	 & 44.3(7.1)  	 & 90(24)     	 & 96(29)                 	 & 99(22)     	 & 31.8(3.7) \\
$\cdots$		    	 & Clear 	 & 41.9(7.2)  	 & 67(24)     	 & 74(25)                 	 & 78(21)     	 & 29.0(5.4) \\
2020 Feb 20$-$26    	 & $B$   	 & 100(52)    	 & 155(46)    	 & 80(100)  	 & 178(48)    	 & 28.7(8.1) \\
$\cdots$		    	 & Clear 	 & 103(53)    	 & 137(47)    	 & 100(100) 	 & 164(49)    	 & 26.5(8.8) \\
2021 Jul 24$-$30    	 & $B$   	 & 46.8(3.7)  	 & 84.5(3.1)  	 & 4(24)                  	 & 96.5(3.2)  	 & 30.5(1.1) \\
$\cdots$		    	 & Clear 	 & 49.2(3.5)  	 & 61.3(2.6)  	 & 20(96)                 	 & 78.6(3.0)  	 & 25.6(1.2) \\
2021 Aug 14$-$18    	 & $B$   	 & 16(13)     	 & 98.92(26)  	 & 90(55)                 	 & 99.4(2.0)  	 & 40.5(3.5) \\
$\cdots$		    	 & Clear 	 & 20(13)     	 & 76.75(61)  	 & 60(52)                 	 & 78.2(3.2)  	 & 37.9(4.5) \\
2021 Sep 14$-$Oct 5 	 & Clear 	 & 36($-$)	 &  69($-$)	 & 19($-$)	 		&  78($-$)	 	& 31($-$) \\
\hline
Time Average                	 & $U$   	 & 126.8(1.0) 	 & 125.2(2.5) 	 & 71(13)                 	 & 178.2(1.9) 	 & 22.31(30) \\
Time Average                	 & $B$   	 & 44(11)     	 & 83.7(9.4)  	 & 46(43)                 	 & 94.0(9.7)  	 & 31.1(3.2) \\
Time Average                	 & $V$   	 & 12.5(1.1)  	 & 42.94(96)  	 & 80.6(4.7)              	 & 44.69(97)  	 & 36.92(68) \\
Time Average                	 & Clear 	 & 36(13)     	 & 63.9(4.7)  	 & 113.2(3.4)             	 & 72.5(7.8)  	 & 30.2(4.7)   
\label{tab:licktp}
\enddata
\end{deluxetable*}

\begin{deluxetable*}{ccccccc}
\tabletypesize{\small}
\tablecaption{Lick 1-m Telescope Polarization}
\tablewidth{0pt}
\tablehead{
\colhead{UT Date Range}	& \colhead{Band}	& \colhead{$q$}	& \colhead{$u$}	& \colhead{$v$}		& \colhead{$p$}	& \colhead{$\Theta$} \\
 					 & & (ppm) 			& (ppm)			& (ppm)			& (ppm)			& ($^\circ$)}
\startdata
2011 Apr 22         	 & $U$ 	 & 868.6(9.1) 	 & 234(12)    	 & 73(19)     	 & 899.5(9.3) 	 & 7.54(38)   \\
$\cdots$	         	 & $B$ 	 & 581(12)    	 & 68(13)     	 & 8(10)      	 & 585(12)    	 & 3.35(64)   \\
$\cdots$	         	 & $V$ 	 & 453(11)    	 & 38(13)     	 & 49(18)     	 & 455(12)    	 & 2.37(82)   \\
2011 May 24$-$27    	 & $B$ 	 & 508(40)    	 & 83(46)     	 & 73(49)     	 & 513(40)    	 & 4.6(2.5)   \\
2011 Jul 11$-$14    	 & $B$ 	 & 503.3(6.2) 	 & 130.1(9.7) 	 & 53(14)     	 & 519.7(6.5) 	 & 7.25(52)   \\
2011 Oct 8$-$25     	 & $U$ 	 & 706(76)    	 & 275(85)    	 & 111(56)    	 & 753(77)    	 & 10.6(3.2)  \\
$\cdots$	     		 & $B$ 	 & 399(77)    	 & 98(86)     	 & 13(56)     	 & 402(77)    	 & 6.9(6.1)   \\
$\cdots$		     	 & $V$ 	 & 275(77)    	 & 73(86)     	 & 23(35)     	 & 272(78)    	 & 7.4(8.9)   \\
2012 Mar 23         	 & $B$ 	 & 404(43)    	 & 167(28)    	 & 190(110)   	 & 436(41)    	 & 11.2(2.0)  \\
$\cdots$		        	 & $V$ 	 & 279(44)    	 & 141(28)    	 & 160(100)   	 & 311(41)    	 & 13.4(3.0)  \\
2012 May 15$-$26    	 & $B$ 	 & 566(33)    	 & 78(27)     	 & 59(11)     	 & 571(33)    	 & 3.9(1.3)   \\
2012 Jul 10$-$26    	 & $B$ 	 & 606(19)    	 & 72(30)     	 & 17(38)     	 & 610(19)    	 & 3.4(1.4)   \\
2012 Aug 8$-$24     	 & $B$ 	 & 499(36)    	 & $-$94(23)  	 & 56(14)     	 & 508(35)    	 & 174.7(1.4) \\
2012 Sep 11$-$20    	 & $B$ 	 & 494(23)    	 & $-$110(35) 	 & 27(25)     	 & 505(24)    	 & 173.7(2.0) \\
2012 Oct 26$-$29    	 & $U$ 	 & 914(39)    	 & 253(62)    	 & 193(20)    	 & 947(41)    	 & 7.7(1.8)   \\
$\cdots$		    	 & $B$ 	 & 608(39)    	 & 77(62)     	 & 100(20)    	 & 610(40)    	 & 3.6(2.9)   \\
$\cdots$		    	 & $V$ 	 & 487(40)    	 & 49(62)     	 & 78(18)     	 & 486(40)    	 & 2.9(3.6)   \\
2013 Apr 7$-$9      	 & $B$ 	 & 556.7(5.6) 	 & 52.0(7.4)  	 & 54(15)     	 & 559.1(5.7) 	 & 2.67(38)   \\
2013 May 22$-$Jun 1 	 & $B$ 	 & 586(51)    	 & 69(51)     	 & 69(34)     	 & 588(51)    	 & 3.4(2.5)   \\
2013 Jun 2$-$28     	 & $B$ 	 & 474(36)    	 & 12(23)     	 & 70(58)     	 & 474(36)    	 & 0.7(1.4)   \\
2013 Aug 13$-$15    	 & $B$ 	 & 595(31)    	 & $-$37(47)  	 & 90(23)     	 & 594(31)    	 & 178.2(2.3) \\
2014 Jan 22$-$27    	 & $U$ 	 & 912(31)    	 & 267(32)    	 & 122(13)    	 & 950(31)    	 & 8.15(97)   \\
$\cdots$		   	 & $B$ 	 & 607(31)    	 & 91(32)     	 & 28(13)     	 & 613(31)    	 & 4.3(1.5)   \\
$\cdots$		    	 & $V$ 	 & 486(31)    	 & 61(33)     	 & 30(13)     	 & 488(31)    	 & 3.6(1.9)   \\
2014 May 24$-$30    	 & $U$ 	 & 907(56)    	 & 225(46)    	 & 115(11)    	 & 933(55)    	 & 7.0(1.4)   \\
$\cdots$		    	 & $B$ 	 & 601(56)    	 & 48(46)     	 & 19(20)     	 & 601(56)    	 & 2.3(2.2)   \\
$\cdots$		    	 & $V$ 	 & 478(56)    	 & 20(46)     	 & 51(57)     	 & 477(56)    	 & 1.2(2.8)   \\
2014 Jun 17$-$20    	 & $U$ 	 & 887(27)    	 & 263(32)    	 & 157.4(6.8) 	 & 925(28)    	 & 8.25(98)   \\
$\cdots$		    	 & $B$ 	 & 582(27)    	 & 86(32)     	 & 66.0(7.0)  	 & 588(27)    	 & 4.2(1.6)   \\
$\cdots$		    	 & $V$ 	 & 460(27)    	 & 58(32)     	 & 68.4(9.2)  	 & 463(27)    	 & 3.6(2.0)   \\
2014 Jul 24$-$Aug 1 	 & $B$ 	 & 697(59)    	 & 20(92)     	 & 3(40)      	 & 691(59)    	 & 0.8(3.8)   \\
2014 Aug 8$-$28     	 & $B$ 	 & 734(72)    	 & 105(63)    	 & 30(16)     	 & 739(72)    	 & 4.1(2.5)   \\
2014 Sep 3$-$28     	 & $B$ 	 & 560(490)   	 & 250(550)   	 & 50(110)    	 & 440(500)   	 & 12(32)     \\
$\cdots$		     	 & $V$ 	 & 440(490)   	 & 200(550)   	 & 70(140)    	 & 320(500)   	 & 12(46)     \\
2014 Oct 11$-$27    	 & $B$ 	 & 546(14)    	 & 69(21)     	 & 54.0(1.9)  	 & 550(14)    	 & 3.6(1.1)   \\
2014 Nov 7$-$8      	 & $B$ 	 & 574(16)    	 & 72(10)     	 & 55.4(2.2)  	 & 579(16)    	 & 3.56(52)   \\
\hline
Time Average        	 & $U$ 	 & 894(22)    	 & 251(16)    	 & 140(28)    	 & 928(21)    	 & 7.86(51)   \\
Time Average        	 & $B$ 	 & 535(50)    	 & 95(68)     	 & 53.8(9.3)  	 & 539(51)    	 & 5.0(3.6)   \\
Time Average        	 & $V$ 	 & 387(29)    	 & 190(45)    	 & 57.4(6.3)  	 & 429(32)    	 & 13.1(2.8)    
\label{tab:licktp1}
\enddata
\end{deluxetable*}

\begin{deluxetable*}{ccccccc}
\tabletypesize{\small}
\tablecaption{Lick 3-m/1-m Telescope Linear Polarization Ratios}
\tablewidth{0pt}
\tablehead{
\colhead{Telescope, Band}	& \colhead{3-m, $U$}	& \colhead{3-m, $B$}	& \colhead{3-m, $V$}	& \colhead{1-m, $U$}	& \colhead{1-m, $B$}	& \colhead{1-m, $V$}}
\startdata
3-m, $U$ 	 & 1      	 & 1.90(20)  	 & 3.987(96) 	 & 0.1919(48) 	 & 0.331(31)  	 & 0.415(32)  \\
3-m, $B$ 	 & 0.527(55)  	 & 1     	 & 2.10(22)  	 & 0.101(11)  	 & 0.174(24)  	 & 0.219(28)  \\
3-m, $V$ 	 & 0.2508(60) 	 & 0.476(51) 	 & 1     	 & 0.0481(15) 	 & 0.0830(80) 	 & 0.1041(82) \\
1-m, $U$ 	 & 5.21(13)   	 & 9.9(1.1)  	 & 20.77(65) 	 & 1      	 & 1.72(17)   	 & 2.16(17)   \\
1-m, $B$ 	 & 3.02(29)   	 & 5.73(81)  	 & 12.1(1.2) 	 & 0.580(56)  	 & 1      	 & 1.25(15)   \\
1-m, $V$ 	 & 2.41(18)   	 & 4.57(59)  	 & 9.61(76)  	 & 0.463(37)  	 & 0.797(96)  	 & 1        
\label{tab:licktplinratios}
\enddata
\end{deluxetable*}

\begin{deluxetable*}{ccccccc}
\tabletypesize{\small}
\tablecaption{Lick 3-m/1-m Telescope Circular Polarization Ratios}
\tablewidth{0pt}
\tablehead{
\colhead{Telescope, Band}	& \colhead{3-m, $U$}	& \colhead{3-m, $B$}	& \colhead{3-m, $V$}	& \colhead{1-m, $U$}	& \colhead{1-m, $B$}	& \colhead{1-m, $V$}}
\startdata
3-m, $U$ 	 & 1    	 & 1.6(2.4) 	 & 0.88(17)  	 & 0.51(14)  	 & 1.32(34) 	 & 1.24(26) \\
3-m, $B$ 	 & 0.64(62) 	 & 1    	 & 0.57(53)  	 & 0.33(32)  	 & 0.85(82) 	 & 0.80(75) \\
3-m, $V$ 	 & 1.13(22) 	 & 1.8(2.7) 	 & 1     	 & 0.58(13)  	 & 1.50(28) 	 & 1.41(18) \\
1-m, $U$ 	 & 1.96(54) 	 & 3.0(4.8) 	 & 1.73(37)  	 & 1     	 & 2.60(71) 	 & 2.44(57) \\
1-m, $B$ 	 & 0.76(19) 	 & 1.2(1.8) 	 & 0.67(12)  	 & 0.39(11)  	 & 1    	 & 0.94(19) \\
1-m, $V$ 	 & 0.80(17) 	 & 1.3(1.9) 	 & 0.712(89) 	 & 0.410(98) 	 & 1.07(22) 	 & 1      
\label{tab:licktpcircratios}
\enddata
\end{deluxetable*}

\subsubsection{Dome Shadow Polarization}

\begin{figure}
\centering
\includegraphics[width=0.47\textwidth]{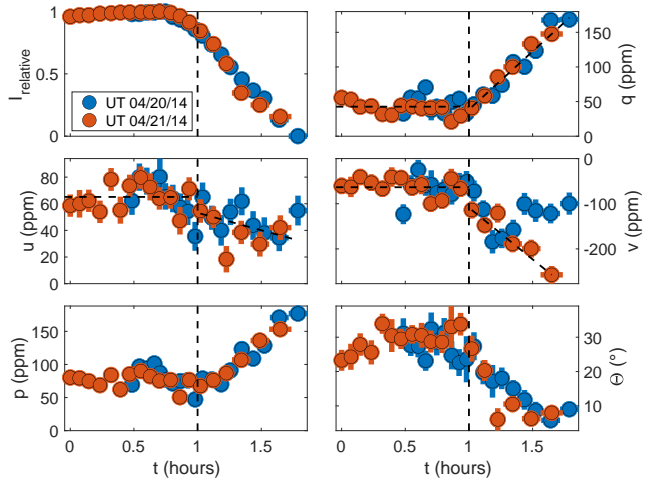}
\caption{Lick 3-m POLISH2 $B$ band measurements of the weakly polarized HR 4295 as it is occulted by the telescope dome. Telescope polarization is subtracted as discussed in Section \ref{sec:lick}. Dashed, vertical lines indicate when dome occultation begins to be apparent in Stokes $q$.}
\label{fig:tpoccult}
\end{figure}

We illustrate the nightly accuracy of POLISH2 observations by allowing the Lick 3-m telescope to track the bright, weakly polarized HR 4295 with dome tracking powered off on two successive nights (UT April 20 and 21, 2014). As the telescope tracks into the dome, the dome shadow sweeps across the primary and secondary mirrors until the star is completely occulted. This generates a time-dependent asymmetry to the illumination of the telescope mirrors, which manifests as a change in the measured polarization of the star (Figure \ref{fig:tpoccult}). This trend is most apparent in Stokes $q$ and $v$ as a linear increase in polarization. In Figure \ref{fig:tpoccult}, we find that Stokes $q$, the strongest, most repeatable trend, departs from a constant value (vertical, dashed lines) when total intensity has been reduced by $100\% - I_\text{relative} \sim 15\%$. Both linear and circular polarization increase by $\sim 100$ ppm (0.01\%) in a repeatable manner from night to night (Table \ref{tab:shadow}), which demonstrates that POLISH2 accuracy is sufficient to search for scattered light modulation of order 10 ppm in exoplanet systems. Interestingly, the slope of circular polarization change, $\dot{v} \sim -230$ ppm/h, is significantly larger in absolute value than that for linear polarization, $\dot{q} \sim 170$ ppm/h. For as-yet unknown reasons, the circular polarization trends from the two nights diverge just before the star is fully occulted.

\begin{figure}
\centering
\includegraphics[width=0.47\textwidth]{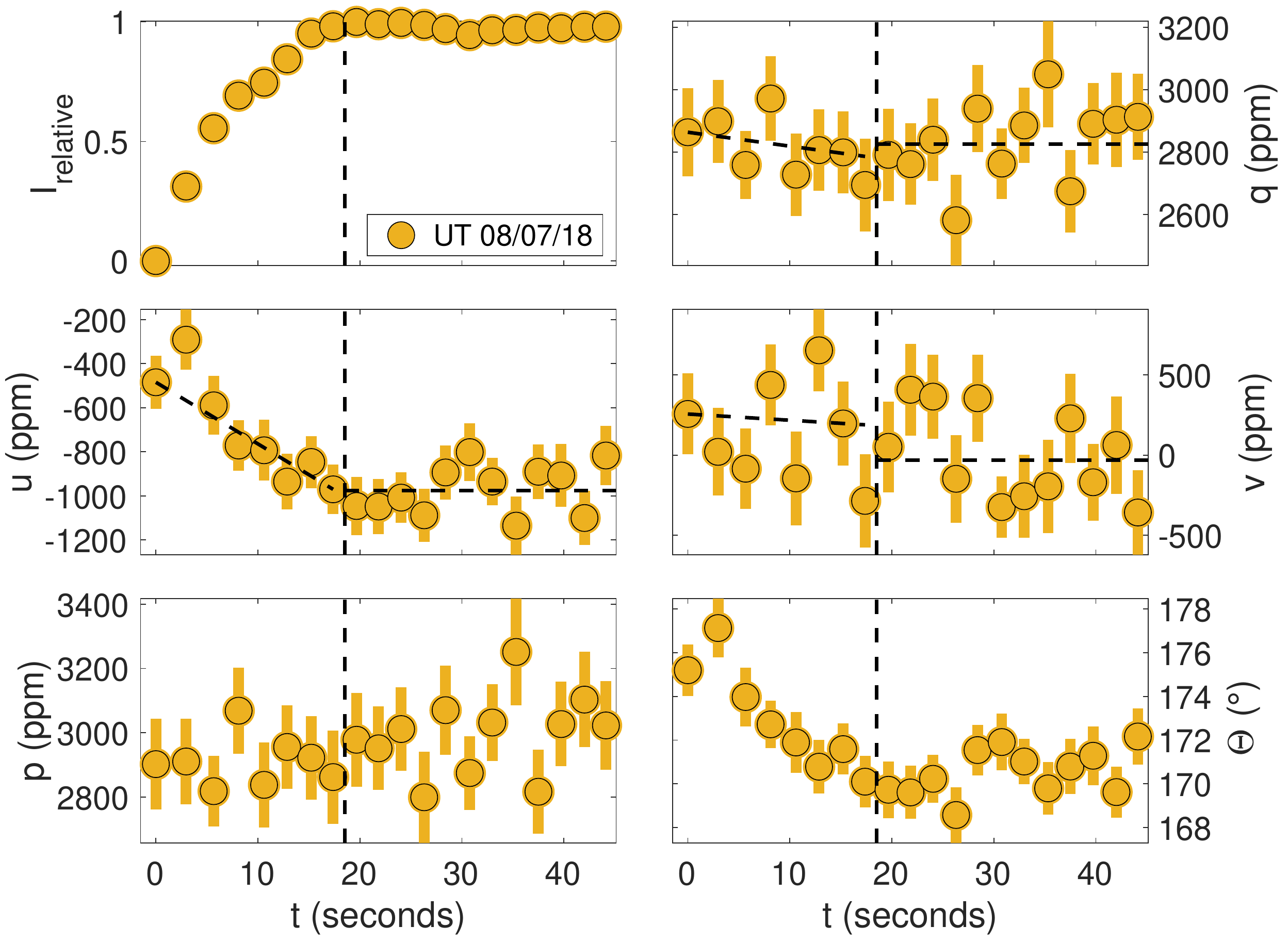}
\caption{The same as Figure \ref{fig:tpoccult} but for Gemini North POLISH2 $B$ band measurements of the weakly polarized HR 8585. Telescope polarization is not subtracted, and its large, $\sim 0.3\%$ value is manifested in the mean values of $q$, $u$, and $p$.}
\label{fig:tpoccultgem}
\end{figure}

This experiment was repeated in reverse at Gemini North on UT August 7, 2018. Here, integrations were started after the telescope had slewed to the bright, weakly polarized HR 8585 but while the dome was still moving to position (Figure \ref{fig:tpoccultgem}). This experiment lasted $\sim 45$ seconds, compared to the $\sim 2$ hour duration of the Lick 3-m experiment on each night, so the slopes of the polarization trends are significantly larger. In Figure \ref{fig:tpoccultgem}, we determine the time at which Stokes $u$, the strongest trend, departs from a constant value (vertical, dashed lines). At this point, total intensity has been reduced by only $100\% - I_\text{relative} \sim 1\%$, compared to $\sim 15\%$ at Lick, which indicates that only a slight dome shadow is required to detect polarization change of the highly polarized Gemini North mirrors. Indeed, while polarization trends $\Delta q$ and $\Delta u$ at Lick and Gemini appear qualitatively similar, respectively, the dramatically larger telescope polarization at Gemini appears to cause a correspondingly larger amplitude of dome shadow polarization.

Dome shadow polarization during our Gemini experiment actually decreases the absolute value of Stokes $u$ telescope polarization, because the dome shadow adds a polarization component of Stokes $+u$ that partially counteracts the large, intrinsic, Stokes $-u$ component measured at the time of the experiment. Indeed, telescope polarization at Gemini North is so large that covering most of the telescope mirrors actually decreases the absolute value of telescope polarization. This may suggest that the cause of the anomalously large Gemini North telescope polarization is localized on the surface of one of the mirrors instead of being uniform in nature.

\begin{deluxetable*}{cccccccccc}
\tabletypesize{\footnotesize}
\tablecaption{Dome Shadow Polarization}
\tablewidth{0pt}
\tablehead{
\colhead{Telescope} & \colhead{UT Date} & \colhead{$\Delta q$} & \colhead{$\Delta u$} & \colhead{$\Delta p$} & \colhead{$\Delta v$} & \colhead{$\dot{q}$} & \colhead{$\dot{u}$} & \colhead{$\dot{v}$} \\
 & & (ppm) & (ppm) & (ppm)	& (ppm) & (ppm/h) & (ppm/h) & (ppm/h)}
\startdata
Lick			& 2014 Apr 20$-$21 	& 128.2(8.7)	& $-33(12)$	& 131.8(8.9)	& $-194(29)$	& 168(11)			& $-26(14)$		& $-229(43)$ \\
Gemini	& 2018 Aug 7		& 40(110)		& 490(140)	& 480(140)	& 290(390)	& $-16(22) \times 10^3$	& $-101(28) \times 10^3$	& $-14(86) \times 10^3$
\label{tab:shadow}
\enddata
\end{deluxetable*}

\section{POLISH2 Polarization Database}
\label{sec:database}

The POLISH2 polarization database is split between weakly polarized targets ($\overline{p} < 0.1\%$, Table \ref{tab:lopoldatabase}, units of ppm) and strongly polarized targets ($\overline{p} \geq 0.1\%$, Table \ref{tab:hipoldatabase}, units of percent). Data were obtained from April 2011 to September 2021 at both the Lick 3-m and 1-m. Gemini North 8-m observations are also included. Linear polarization values from catalogs, if applicable, are listed in columns $p_{\text{cat}}$ and $\Theta_{\text{cat}}$ with references denoted.

An estimate of intrinsic variability of targets observed over at least two nights is given in Table \ref{tab:vardatabase}. Intrinsic variability in each Stokes parameter is calculated by the following, with $q_\text{var}$ as an example and $\sigma_\text{wv}$ defined in Equation \ref{eq:std}:
\begin{eqnarray}
q_\text{var} & = \operatorname{Re} \left[ \sqrt{\sigma_{\text{wv}_q}^2 - \texttt{median}(\sigma_q)^2} \right].
\end{eqnarray}

\noindent We provide context for the following objects.

\startlongtable
\begin{deluxetable*}{ccccccccccccccccc}
\tabletypesize{\tiny}
\tablecaption{POLISH2 Database of Weakly Polarized Targets ($\overline{p} < 0.1\%$)}
\tablewidth{0pt}
\tablehead{
\colhead{Tel.} & \colhead{Object} & \colhead{UT Date Range} & \colhead{Band} & \colhead{$\overline{q}$ (ppm)} & \colhead{$\overline{u}$ (ppm)} & \colhead{$\overline{v}$ (ppm)} & \colhead{$\overline{p}$ (ppm)} & \colhead{$\overline{\Theta}$ ($^\circ$)} & \colhead{$p_{\text{cat}}$ (ppm)} & \colhead{$\Theta_{\text{cat}}$ ($^\circ$)}}
\startdata
3-m	& $\beta$ Oph	& 2013 May 26			& $B$             	 & 21(15)        	 & 14(14)        	 & $-$18(29)     	 & 22(14)     	 & 17(18)     	 & 0.618(44)    	 & 35.3(1.4)    \tablenotemark{f} \\
3-m	& $\epsilon$ UMa	& 2014 Mar 8$-$19			& $B$      	 & 5.9(9.9)      	 & $-$9.5(2.9)   	 & 8(35)         	 & 8.5(5.8)   	 & 151(26)    	 & 100(1200)    	 & 143(80)      \tablenotemark{c} \\
1-m	& Aldebaran	& 2011 Oct 8$-$22			& $B$           	 & $-$8(35)      	 & 78(23)        	 & 26(84)        	 & 70(23)     	 & 48(13)     	 & $-$     	 & $-$     \tablenotemark{ } \\
3-m	& Algol		& 2013 Sep 11$-$2015 Sep 29		& $B$     	 & $-$3(27)      	 & $-$91(69)     	 & $-$2(42)      	 & 87(69)     	 & 134(11)    	 & $-$21.0(9.0) 	 & $-$0.0070(10)\tablenotemark{e} \\
1-m	& Algol		& 2014 Jan 22$-$Oct 12		& $B$          	 & 25(69)        	 & $-$225(91)    	 & 50(300)       	 & 216(91)    	 & 138.2(9.4) 	 & $-$21.0(9.0) 	 & $-$0.0070(10)\tablenotemark{e} \\
1-m	& Algol		& 2014 Sep 4			& $V$                   	 & 276(32)       	 & $-$114(30)    	 & 904(79)       	 & 297(31)    	 & 168.8(2.9) 	 & $-$21.0(9.0) 	 & $-$0.0070(10)\tablenotemark{e} \\
1-m	& Castor		& 2011 Oct 8			& $B$                  	 & $-$330(17)    	 & $-$28(15)     	 & $-$571(40)    	 & 331(17)    	 & 92.4(1.3)  	 & $-$     	 & $-$     \tablenotemark{ } \\
3-m	& HD 25621	& 2015 Feb 11			& Clear              	 & 25(42)        	 & 76(33)        	 & $-$110(36)    	 & 70(34)     	 & 36(16)     	 & $-$     	 & $-$     \tablenotemark{ } \\
\enddata
\tablenotetext{a}{\cite{Michalsky1976}}
\tablenotetext{b}{\cite{Schmidt1992}}
\tablenotetext{c}{\cite{Heiles2000}}
\tablenotetext{d}{\cite{Weitenbeck2004}}
\tablenotetext{e}{\cite{Elias2008}}
\tablenotetext{f}{\cite{Bailey2010}}
\tablenotetext{g}{\cite{Ababakr2016}}
\tablenotetext{h}{\cite{Cotton2017}}
\tablenotetext{i}{\cite{Bailey2020}}
\end{deluxetable*}

\startlongtable
\begin{deluxetable*}{ccccccccccccccccc}
\tabletypesize{\tiny}
\tablecaption{POLISH2 Database of Strongly Polarized Targets ($\overline{p} \geq 0.1\%$)}
\tablewidth{0pt}
\tablehead{
\colhead{Tel.} & \colhead{Object} & \colhead{UT Date Range} & \colhead{Band} & \colhead{$\overline{q}$ (\%)} & \colhead{$\overline{u}$ (\%)} & \colhead{$\overline{v}$ (\%)} & \colhead{$\overline{p}$ (\%)} & \colhead{$\overline{\Theta}$ ($^\circ$)} & \colhead{$p_{\text{cat}}$ (\%)} & \colhead{$\Theta_{\text{cat}}$ ($^\circ$)}}
\startdata
3-m	& 10 Sge		& 2011 Jul 21$-$22			& $B$                      	 & 0.6075(27)     	 & 0.34143(26)    	 & $-$0.02162(89) 	 & 0.6968(23)  	 & 14.669(55)   	 & 0.6890(90) 	 & 15.50(40) \tablenotemark{c} \\
1-m	& 10 Sge		& 2011 Jul 11			& $B$                           	 & 0.563(13)      	 & 0.394(14)      	 & $-$0.025(27)   	 & 0.687(13)   	 & 17.48(56)    	 & 0.6890(90) 	 & 15.50(40) \tablenotemark{c} \\
3-m	& 49 Tau		& 2011 Dec 5			& $B$                            	 & 0.2207(19)     	 & 0.2834(17)     	 & 0.0156(36)     	 & 0.3592(18)  	 & 26.04(15)    	 & 0.37(20)   	 & 15(15)    \tablenotemark{c} \\
3-m	& 53 Per		& 2014 Jan 4$-$Mar 19		& $B$                    	 & 0.8355(53)     	 & $-$0.1348(37)  	 & $-$0.0078(60)  	 & 0.8463(53)  	 & 175.42(13)   	 & 1.15(20)   	 & 172.0(5.0)\tablenotemark{c} \\
8-m	& 53 Per		& 2016 Nov 12			& Clear                         	 & 0.7418(25)     	 & $-$0.1282(13)  	 & 0.00603(89)    	 & 0.7528(25)  	 & 175.098(53)  	 & 1.15(20)   	 & 172.0(5.0)\tablenotemark{c} \\
3-m	& 53 Per		& 2014 Mar 17$-$2020 Feb 20		& Clear            	 & 0.90(25)       	 & $-$0.053(33)   	 & 0.0061(37)     	 & 0.90(25)    	 & 178.3(1.2)   	 & 1.15(20)   	 & 172.0(5.0)\tablenotemark{c} \\
3-m	& 55 Cyg		& 2014 Apr 20$-$21			& $U$                      	 & 3.164(52)      	 & 0.2090(36)     	 & $-$0.127(16)   	 & 3.171(52)   	 & 1.890(45)    	 & 2.877(90)  	 & 2.50(90)  \tablenotemark{c} \\
3-m	& 55 Cyg		& 2011 Aug 17$-$2019 Oct 25		& $B$              	 & 2.729(67)      	 & 0.133(36)      	 & $-$0.036(13)   	 & 2.732(67)   	 & 1.40(38)     	 & 2.877(90)  	 & 2.50(90)  \tablenotemark{c} \\
\enddata
\tablenotetext{a}{\cite{Michalsky1976}}
\tablenotetext{b}{\cite{Schmidt1992}}
\tablenotetext{c}{\cite{Heiles2000}}
\tablenotetext{d}{\cite{Weitenbeck2004}}
\tablenotetext{e}{\cite{Elias2008}}
\tablenotetext{f}{\cite{Bailey2010}}
\tablenotetext{g}{\cite{Ababakr2016}}
\tablenotetext{h}{\cite{Cotton2017}}
\tablenotetext{i}{\cite{Bailey2020}}
\end{deluxetable*}

\startlongtable
\begin{deluxetable*}{ccccccccccccccccc}
\tabletypesize{\tiny}
\tablecaption{Variability of POLISH2 Targets}
\tablewidth{0pt}
\tablehead{
\colhead{Tel.} & \colhead{Object} & \colhead{UT Date Range} & \colhead{Band} & \colhead{$n_\text{nights}$} & \colhead{$q_\text{var}$ (ppm)} & \colhead{$u_\text{var}$ (ppm)} & \colhead{$v_\text{var}$ (ppm)}}
\startdata
3-m	& 10 Sge		& 2011 Jul 21$-$22			& $B$                     	&  2	&     4	&    0	&    0 \\
3-m	& 53 Per		& 2014 Jan 4$-$Mar 19		& $B$                   	& 14	&   183	&   34	&   72 \\
3-m	& 53 Per		& 2014 Mar 17$-$2020 Feb 20		& Clear           	&  3	&  5909	&  371	&   81 \\
3-m	& 55 Cyg		& 2014 Apr 20$-$21			& $U$                     	&  2	&   496	&    0	&    0 \\
3-m	& 55 Cyg		& 2011 Aug 17$-$2019 Oct 25		& $B$             	&  6	&  1010	&  355	&  106 \\
3-m	& 55 Cyg		& 2014 Apr 20$-$21			& $V$                     	&  2	&   350	&    0	&   44 \\
3-m	& 55 Cyg		& 2019 Oct 23$-$25			& Clear                   	&  3	&   773	&  140	&    0 \\
1-m	& 66 Oph		& 2011 Jul 11$-$2012 Mar 23		& $B$             	&  2	&    58	&    0	&    0 \\
\enddata
\end{deluxetable*}

\subsection{Circular Polarization}
\label{sec:circobj}

\begin{figure*}
\centering
\includegraphics[width=1\textwidth]{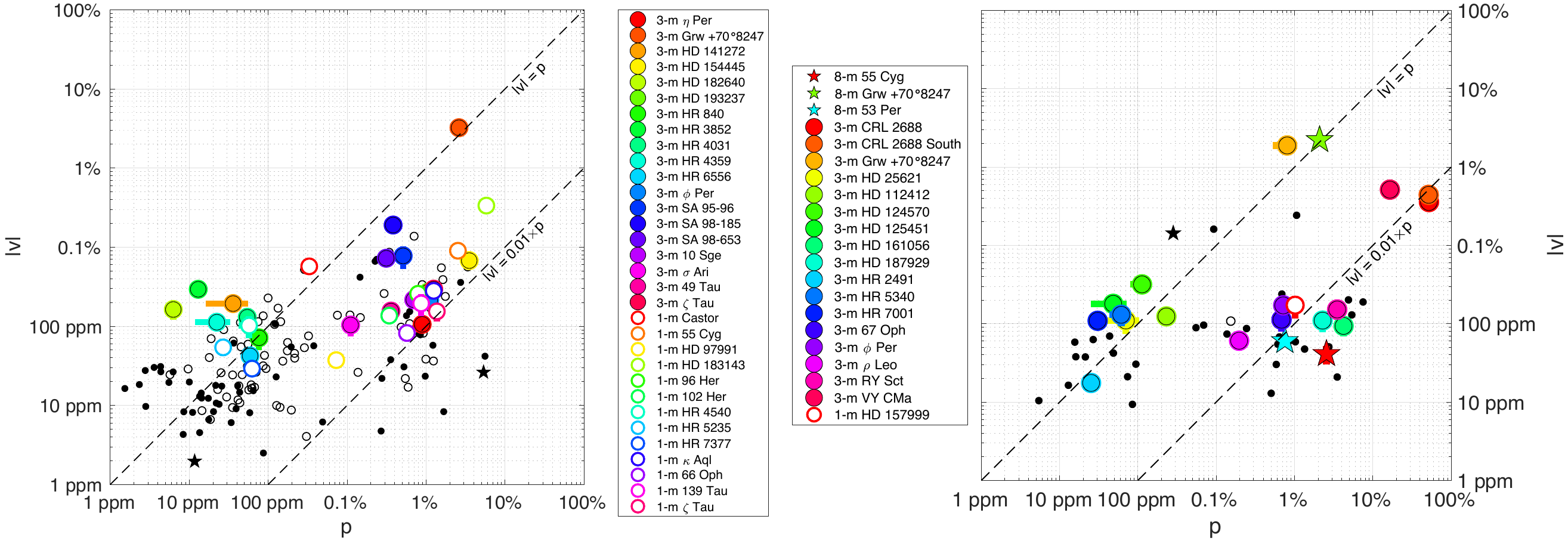}
\caption{\textit{Left:} Absolute value of $B$ band circular polarization $|v|$ versus degree of linear polarization $p$ obtained with POLISH2 at Gemini North (open stars), the Lick 3-m (filled circles), and the Lick 1-m (open circles). Stars with circular polarization detected with at least $3 \sigma$ confidence are plotted in large, colored stars or circles and labeled in the legend. Dashed lines indicate circular polarization scaling with degree of linear polarization with 1:1 and 1:100 ratios. \textit{Right:} Same as the left panel but for the clear POLISH2 bandpass.}
\label{fig:circBC}
\end{figure*}

Figure \ref{fig:circBC} compares POLISH2 measurements of linear and circular polarization for individual objects observed at Gemini North, the Lick 3-m, and the Lick 1-m. Table \ref{tab:circdatabase} lists objects with circular polarization detected with at least $|\overline{v}|/\sigma_v = 3 \sigma$ confidence. These objects are sorted by time-averaged degree of linear polarization $\overline{p}$. Observations \citep[e.g.,][]{Kemp1972HD154445} and theory \citep[e.g.,][]{Martin1974} show that variations in the alignment of ISM dust grains along the line of sight to a polarized object may convert linear to circular polarization with efficiency of order $|\overline{v}|/\overline{p} \sim 0.01$. Our measurements in Table \ref{tab:circdatabase} tend to have similar efficiencies save for Grw $+70^\circ8247$.

\startlongtable
\begin{deluxetable*}{ccccccccccccccccc}
\tabletypesize{\tiny}
\tablecaption{POLISH2 Detections of Circular Polarization}
\tablewidth{0pt}
\tablehead{
\colhead{Tel.} & \colhead{Object} & \colhead{Band} & \colhead{$\overline{v}$ (\%)} & \colhead{$|\overline{v}|/\sigma_v$} & \colhead{$\overline{p}$ (\%)} & \colhead{$|\overline{v}|/\overline{p}$}}
\startdata
3-m	& HD 182640          	& $B$ 	 & $-$0.0163(41)  	& 4.0	 & 0.0006(20)   	& 30(380)     \\
3-m	& HR 3852            	& $B$ 	 & $-$0.0293(57)  	& 5.1	 & 0.0013(28)   	& 20(150)     \\
3-m	& HR 4359            	& $B$ 	 & 0.0112(22)     	& 5.2	 & 0.0022(10)   	& 5.0(3.0)    \\
3-m	& HR 2491            	& $C$ 	 & 0.001753(67)   	&  26	 & 0.00252(10)  	& 0.696(38)   \\
1-m	& HR 5235            	& $B$ 	 & $-$0.0054(12)  	& 4.5	 & 0.0027(37)   	& 2.0(6.1)    \\
3-m	& HR 7001            	& $C$ 	 & 0.0109(13)     	& 8.7	 & 0.003057(75) 	& 3.57(42)    \\
3-m	& HD 141272          	& $B$ 	 & 0.0194(40)     	& 4.8	 & 0.0036(20)   	& 5.4(4.0)    \\
3-m	& HD 125451          	& $C$ 	 & $-$0.0180(24)  	& 7.4	 & 0.0048(23)   	& 3.7(2.2)    \\
3-m	& HR 4031            	& $B$ 	 & $-$0.0130(15)  	& 8.6	 & 0.00543(75)  	& 2.39(44)    \\
1-m	& HR 4540            	& $B$ 	 & $-$0.0102(32)  	& 3.2	 & 0.0057(15)   	& 1.78(77)    \\
3-m	& HR 6556            	& $B$ 	 & 0.00422(67)    	& 6.3	 & 0.00590(32)  	& 0.72(12)    \\
3-m	& HR 5340            	& $C$ 	 & $-$0.0130(37)  	& 3.5	 & 0.0061(18)   	& 2.13(92)    \\
1-m	& HR 7377            	& $B$ 	 & 0.00292(32)    	& 9.1	 & 0.0062(14)   	& 0.47(12)    \\
3-m	& HD 25621           	& $C$ 	 & $-$0.0110(36)  	& 3.1	 & 0.0070(34)   	& 1.6(1.1)    \\
3-m	& HR 840             	& $B$ 	 & 0.0071(22)     	& 3.3	 & 0.0076(12)   	& 0.94(33)    \\
3-m	& HD 124570          	& $C$ 	 & $-$0.0318(29)  	&  11	 & 0.0113(34)   	& 2.82(95)    \\
3-m	& HD 112412          	& $C$ 	 & $-$0.01240(92) 	&  14	 & 0.02308(48)  	& 0.537(41)   \\
1-m	& Castor             	& $B$ 	 & $-$0.0571(40)  	&  14	 & 0.0331(17)   	& 1.73(15)    \\
1-m	& HD 97991           	& $B$ 	 & 0.00373(20)    	&  19	 & 0.0727(31)   	& 0.0513(35)  \\
3-m	& $\sigma$ Ari       	& $B$ 	 & $-$0.0104(30)  	& 3.5	 & 0.1107(16)   	& 0.094(27)   \\
3-m	& $\rho$ Leo         	& $C$ 	 & $-$0.00607(90) 	& 6.8	 & 0.1959(12)   	& 0.0310(46)  \\
3-m	& SA 98-653          	& $B$ 	 & 0.0724(76)     	& 9.5	 & 0.312(17)    	& 0.232(27)   \\
1-m	& 102 Her            	& $B$ 	 & $-$0.0136(29)  	& 4.6	 & 0.3410(84)   	& 0.0400(87)  \\
3-m	& 49 Tau             	& $B$ 	 & 0.0156(36)     	& 4.3	 & 0.3592(18)   	& 0.043(10)   \\
3-m	& SA 98-185          	& $B$ 	 & 0.190(17)      	&  11	 & 0.3818(38)   	& 0.498(46)   \\
3-m	& SA 95-96           	& $B$ 	 & $-$0.078(25)   	& 3.1	 & 0.511(27)    	& 0.152(49)   \\
1-m	& 66 Oph             	& $B$ 	 & $-$0.00821(18) 	&  47	 & 0.5717(50)   	& 0.01436(33) \\
3-m	& 67 Oph             	& $C$ 	 & 0.0113(33)     	& 3.4	 & 0.6825(37)   	& 0.0165(48)  \\
3-m	& 10 Sge             	& $B$ 	 & $-$0.02162(89) 	&  24	 & 0.6968(23)   	& 0.0310(13)  \\
3-m	& $\phi$ Per         	& $C$ 	 & $-$0.01715(93) 	&  18	 & 0.718(37)    	& 0.0239(18)  \\
8-m	& 53 Per             	& $C$ 	 & 0.00603(89)    	& 6.8	 & 0.7528(25)   	& 0.0080(12)  \\
1-m	& 96 Her             	& $B$ 	 & $-$0.0257(36)  	& 7.1	 & 0.7910(67)   	& 0.0325(46)  \\
3-m	& Grw $+70^\circ$8247	& $C$ 	 & 1.89(18)       	&  10	 & 0.80(27)     	& 2.35(92)    \\
1-m	& 139 Tau            	& $B$ 	 & $-$0.0194(59)  	& 3.3	 & 0.8647(76)   	& 0.0225(68)  \\
3-m	& $\eta$ Per         	& $B$ 	 & 0.0106(16)     	& 6.5	 & 0.8903(60)   	& 0.0119(18)  \\
1-m	& HD 157999          	& $C$ 	 & 0.0172(55)     	& 3.1	 & 1.0182(57)   	& 0.0169(54)  \\
3-m	& HD 193237          	& $B$ 	 & $-$0.0271(31)  	& 8.8	 & 1.061(20)    	& 0.0255(29)  \\
3-m	& $\phi$ Per         	& $B$ 	 & $-$0.0213(48)  	& 4.4	 & 1.111(38)    	& 0.0192(44)  \\
1-m	& $\kappa$ Aql       	& $B$ 	 & $-$0.0280(49)  	& 5.7	 & 1.247(29)    	& 0.0225(40)  \\
3-m	& $\zeta$ Tau        	& $B$ 	 & $-$0.0294(10)  	&  29	 & 1.2473(84)   	& 0.02360(84) \\
1-m	& $\zeta$ Tau        	& $B$ 	 & $-$0.0154(40)  	& 3.9	 & 1.3661(22)   	& 0.0113(29)  \\
8-m	& Grw $+70^\circ$8247	& $C$ 	 & 2.220(16)      	& 136	 & 2.091(15)    	& 1.062(11)   \\
3-m	& HD 187929          	& $C$ 	 & 0.0110(32)     	& 3.4	 & 2.2838(67)   	& 0.0048(14)  \\
1-m	& 55 Cyg             	& $B$ 	 & $-$0.090(16)   	& 5.6	 & 2.5268(88)   	& 0.0356(64)  \\
8-m	& 55 Cyg             	& $C$ 	 & $-$0.0041(11)  	& 3.7	 & 2.557(14)    	& 0.00161(43) \\
3-m	& Grw $+70^\circ$8247	& $B$ 	 & 3.23(14)       	&  23	 & 2.604(72)    	& 1.240(65)   \\
3-m	& RY Sct             	& $C$ 	 & 0.0152(40)     	& 3.8	 & 3.505(77)    	& 0.0043(12)  \\
3-m	& HD 154445          	& $B$ 	 & 0.0676(45)     	&  15	 & 3.5139(26)   	& 0.0192(13)  \\
3-m	& HD 161056          	& $C$ 	 & 0.0094(25)     	& 3.8	 & 4.23(14)     	& 0.00222(59) \\
1-m	& HD 183143          	& $B$ 	 & $-$0.335(64)   	& 5.3	 & 5.787(32)    	& 0.058(11)   \\
3-m	& VY CMa             	& $C$ 	 & 0.5150(72)     	&  71	 & 16.532(82)   	& 0.03115(46) \\
3-m	& CRL 2688 South     	& $C$ 	 & 0.442(14)      	&  31	 & 51.701(24)   	& 0.00856(28) \\
3-m	& CRL 2688           	& $C$ 	 & 0.3581(90)     	&  40	 & 52.08(83)    	& 0.00687(21) \\
\label{tab:circdatabase}
\enddata
\end{deluxetable*}

\textit{Grw $\mathit{+70^\circ8247}$ (LAWD 73)}: While it is not difficult to prepare laboratory sources of significant optical circular polarization, astronomical calibrators with percent-level, broadband circular polarization are rare. The source with the largest known broadband optical circular polarization appears to be the magnetic white dwarf Grw $+70^\circ8247$, whose percent-level circular polarization was discovered by \cite{Kemp1970}. Lying at a Dec of nearly $+71^\circ$, this target is unfortunately not visible to most Southern Hemisphere telescopes. For Northern Hemisphere telescopes such as the Lick 3-m, however, Grw $+70^\circ8247$'s polar location enables it to be observable every month of the year save January. This star hosts a significant dependence of circular polarization with wavelength: $v$ nearly vanishes near 370 nm, increases to $\sim 6\%$ at 450 nm, decreases to $\sim 3\%$ at 520 nm, and remains roughly constant through 675 nm. Circular polarization has varied by up to $\sim 1\%$ from 1972 to 2018, though this is primarily confined to $\lambda < 400$ nm, $\lambda \approx 450$ nm, and $600 < \lambda < 675$ nm \citep{Angel1972Grw, Landstreet1975, Angel1985, Putney1995, Bagnulo2019}. Thus, while Grw $+70^\circ8247$'s strong, spectrally-dependent polarization and temporal variability do not lend themselves to accurate calibration of circular polarization modulation efficiency, it is an excellent on-sky target both to verify that the instrument responds to circular polarization and to calibrate the sign of circular polarization. We also find a strong wavelength dependence in both linear and circular polarization (Table \ref{tab:hipoldatabase}). Note that clear, unfiltered observations of Grw $+70^\circ8247$ obtained at Gemini North and the Lick 3-m utilized blue- and red-sensitive PMTs, respectively. Thus, the large linear polarization obtained at Gemini North with unfiltered, blue-sensitive PMTs is consistent with $B$ band measurements obtained at the Lick 3-m. Curiously, circular polarization obtained at Gemini North is consistent with measurements using unfiltered, red-sensitive PMTs at the Lick 3-m.

\textit{VY CMa}: This red hypergiant, lying at $\sim -26^\circ$ Dec, was discovered by \cite{Gehrels1972} to harbor circular polarization of $v \sim 0.4\%$ in $I$ band with a $1 \sigma$ upper limit of $v \sim 0.15\%$ in $V$ band. Our value of $v \sim 0.5\%$ in an unfiltered, red PMT bandpass (Table \ref{fig:qe}) from October 2019 to February 2020 is similar to the value obtained 48 years prior. However, VY CMa's linear polarization is strongly variable, as \cite{Serkowski2001} measure it to vary from 14.2\% to 8.5\% in $V$ band from March 1973 to February 1976. Fascinatingly, we have discovered that $V$ band linear polarization may be cyclical with a 19-year period or harmonics thereof, as our measurements ($p = 16.532 \pm 0.082\%$, $\Theta = 125.12 \pm 0.32^\circ$) are similar to those from January 2001 ($p = 17.36 \pm 0.05\%$, $\Theta = 124^\circ$). Indeed, while the January 2001 observation is a significant outlier in \cite{Serkowski2001}, it is strikingly close to our measurements given the scale of observed variability. While faint, magnetic white dwarfs have been discovered to harbor significant broadband circular polarization \citep{Bagnulo2020, Berdyugin2022}, we find that VY CMa's brightness, circular polarimetric stability, and moderate Dec strongly advocate for it to be utilized as a circular polarization calibrator redward of $V$ band.

\textit{CRL 2688 (Egg Nebula, RAFGL 2688)}: This protoplanetary nebula appears to harbor strong linear and circular polarization \citep[$p = 47.8 \pm 0.4\%$ and $v = 0.75 \pm 0.18\%$ in $V$ band, ][]{Michalsky1976}. Note that this paper defines negative circular polarization as a clockwise rotation of the electric field vector as seen by an observer facing the source, while we define that to be positive circular polarization after others in the community \citep{Shurcliff1962, Landi2004, Landi2007, Bagnulo2009, Bagnulo2019}. When accounting for this, we measure the same sign of circular polarization for CRL 2688 as \cite{Michalsky1976}. While they show that linear polarization increases significantly toward the red, suggesting the generation of dust grains larger than those typically present in the ISM, circular polarization appears to be constant with wavelength in the optical. Our linear polarization measurements are consistent with \cite{Michalsky1976}, especially given our unfiltered bandpass, but our measurements of circular polarization with $v \sim 0.4\%$ are half those from that paper. Adding to the puzzle, the GASP polarimeter measures $p = 51.0 \pm 0.04\%$ linear polarization similar to POLISH2, but it measures $v = -0.1 \pm 0.05\%$ circular polarization \citep{Collins2013}. We hypothesize that uncorrected, instrumental linear to circular polarization conversion leads to circular polarization offsets between the measurements of \cite{Michalsky1976} and \cite{Collins2013}. We also hypothesize that the GASP results provide a false negative for the presence of circular polarization in CRL 2688.

\textit{OMC-1}: Observations at 2.2 $\mu$m \citep{Lonsdale1980} and 3.45 $\mu$m \citep{Serkowski1973} uncovered percent-level circular polarimetry of the Becklin Neugebauer (BN) object in OMC-1, while $K_n$ band imaging polarimetry discovered up to $-17\%$ circular polarization \citep{Chrysostomou2000} in the SEBN region located $\sim 20$ arcsec southeast of the BN object \citep{Aitken1985}. Peak circular polarization in this region reduces to $-7\%$ in $H$ band and disappears in $J$ band. While Lick 3-m POLISH2 observations demonstrate a non-detection of circular polarization in the BN object ($v = 60 \pm 89$ ppm), we discover circular polarization in the SEBN region with $3.9\sigma$ confidence ($v = -139 \pm 36$ ppm). \cite{Serkowski1973} and \cite{Chrysostomou2000} utilize the opposite sign convention for circular polarization as in this paper, and after converting these to our convention, these observations agree with POLISH2 results demonstrating that the circular polarization in the SEBN region of OMC-1 is negative in sign.

\subsection{Other Objects}
\label{sec:otherobj}

\begin{figure}
\centering
\includegraphics[width=0.47\textwidth]{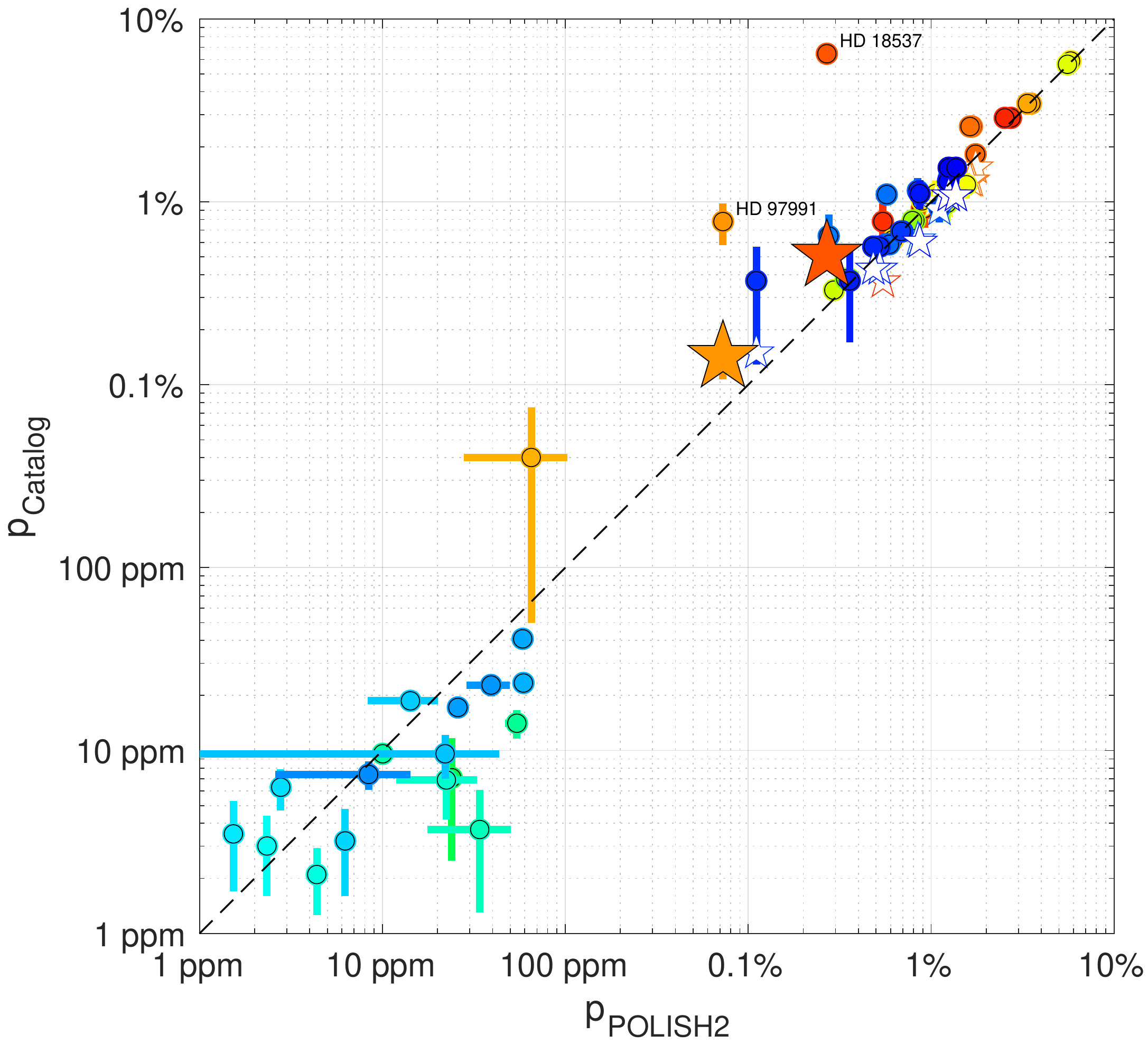}
\caption{Comparison of the measured degree of polarization $p$ for a sample of stars between POLISH2, Starphysics Observatory 0.35-m CGPOL (open stars), and the literature \cite[filled circles]{Heiles2000}. POLISH2 measurements were obtained at both the Lick 3-m and 1-m telescopes. CGPOL measurements of all stars are consistent with POLISH2, which validates POLISH2 results asserting the \cite{Heiles2000} values to be inaccurate for HD 18537 and HD 97991 (large, filled stars).}
\label{fig:catcomp}
\end{figure}

Our polarization results are generally consistent with the catalogs referenced in Tables \ref{tab:lopoldatabase} and \ref{tab:hipoldatabase} \citep{Michalsky1976, Schmidt1992, Heiles2000, Weitenbeck2004, Elias2008, Bailey2010, Ababakr2016, Cotton2017, Bailey2020}. However, Figure \ref{fig:catcomp} shows that we measure degree of linear polarization for HD 18537 and HD 97991 to be more than an order of magnitude lower than the values in the \cite{Heiles2000} catalog. To determine which values are correct, we observed a sample of polarized stars with the CGPOL polarimeter at the Starphysics Observatory 0.35-m \citep{Coleinstpaper2019}. Not only does CGPOL reproduce the POLISH2-measured degree of polarization for this sample, but it also confirms that the \cite{Heiles2000} values for HD 18537 and HD 97991 are inaccurate. Speculation as to the cause of this discrepancy is beyond the scope of this paper.

\section{Conclusion}

We describe our calibration methodology for linear and circular polarimetry with the POLISH2 polarimeter at the Gemini North, Lick Observatory 3-m, and Lick Observatory 1-m telescopes. While telescope-induced polarization at Gemini North has been severe prior to 2021, we demonstrate that self-calibration of science target polarization may accurately remove telescope polarization and uncover intrinsic target polarization at the part-per-million level. Additionally, private communication with Gemini Observatory suggests that telescope polarization may have significantly decreased in 2021. Thus, we expect future polarization measurements with POLISH2 or our next-generation PHALANX (Polarimeter for High Accuracy aLbedos of Asteroids aNd eXoplanets) polarimeters to be sensitive to scattered light from exoplanets at Gemini.

In addition to mitigating telescope polarization at Gemini North, this investigation has halved the variability in Lick 3-m telescope polarization measured by \cite{Wiktorowicz2015_189}. This telescope is demonstrated to harbor intrinsic polarization variability at the level of 10 ppm or below over ten years, which obviates the need to calibrate for potential nightly changes in telescope polarization. At the Lick 1-m, telescope polarization variability is measured to be 50 ppm or less over a timescale of three years. This showcases the utility of relatively small, inexpensive telescopes for the study of relatively bright objects, for which a dense observing cadence may be possible.

We demonstrate the difficulty in obtaining high accuracy, absolute calibration of polarimeters in the laboratory due to the extreme sensitivity of polarization to asymmetry and phase retardance in the laboratory system. Indeed, we measure the linear to circular polarization conversion (``crosstalk") inherent in thin-film polarizers as a pair of them is crossed, and we measure non-zero, intrinsic polarization of cavity blackbodies, integrating spheres, and a variety of light sources. At the Lick 3-m telescope, we demonstrate that rotation of the Cassegrain rotator while continuously acquiring data under a sunlit sky is a powerful technique for measuring and correcting for crosstalk intrinsic to POLISH2. By observing the repeatable change in linear and circular polarization of weakly polarized stars as the Gemini North and Lick 3-m dome slits are allowed to sweep across the telescope mirrors, we demonstrate the accuracy of POLISH2. Such measurements may be made with essentially zero overhead as the telescope and dome slew to a new target, because telescopes tend to arrive at the field faster than their domes.

We tabulate a database of POLISH2 linear and circular polarimetry of a variety of objects, and we caution that essentially every target appears to have measurable variability due to the ISM or intrinsic processes. Few of our targets are shown to have Stokes parameters consistent within error across two or more runs, which suggests that even these targets would show variability given long enough integration time. Thus, we advocate for the standard approach of comparing results to polarization catalogs in aggregate, rather than assuming that catalog values for individual stars are accurate at all times. 

\begin{acknowledgments}

We would like to acknowledge Pushkar Kopparla and Tristan Wolfe for participating in the Gemini POLISH2 commissioning run. We acknowledge Melody Sanchez and Erica Gonzales for their contributions to Lick 1-m POLISH2 observations. SJW acknowledges funding from the NASA SSO (grant NNX17AH84G), XRP (grant 80NSSC19K1578), and Aerospace Technical Investment Programs. SJW and AS acknowledge Polish National Science Centre grant 2017/25/B/ST9/02805.

\end{acknowledgments}

\facilities{Gemini:Gillett (POLISH2), Shane (POLISH2), Nickel (POLISH2), Starphysics (CGPOL)}

\bibliographystyle{apj}
\bibliography{myrefs}
 
 \end{document}